\documentclass[preprint,12pt]{elsarticle}

\usepackage{amssymb}
\usepackage{url}
\usepackage[hidelinks]{hyperref}
\usepackage{soul}
\usepackage[table,xcdraw]{xcolor}
\usepackage{longtable}
\usepackage{lipsum}
\usepackage{lscape} 
\usepackage{booktabs}
\usepackage[utf8]{inputenc}
\usepackage[T1]{fontenc}
\usepackage{lmodern}
\usepackage{microtype}
\usepackage{float}
\usepackage{caption}
\usepackage{blindtext}
\usepackage{siunitx}
\usepackage{amssymb}
\usepackage{colortbl}
\usepackage{multirow}
\usepackage{hhline}
\usepackage{vcell}
\usepackage{mdframed}
\usepackage{fontawesome}
\usepackage{hyperref} 
\usepackage[normalem]{ulem}
\usepackage{tablefootnote}
\usepackage{xcolor}
\usepackage{comment}
\makeatletter
\newcommand\footnoteref[1]{\protected@xdef\@thefnmark{\ref{#1}}\@footnotemark}
\makeatother

\journal{Information and Software Technology}

\begin{document}

\begin{frontmatter}

\title{Zero-Shot Learning for Requirements Classification:\\ an Exploratory Study}

\author[inst1]{Waad Alhoshan}
\affiliation[inst1]{organization={College of Computer and Information Sciences},
            addressline={Imam Mohammad Ibn Saud Islamic University (IMSIU)},
            city={Riyadh},
            postcode = {11564},
            country={Saudi Arabia}}
\author[inst2]{Alessio Ferrari}
\affiliation[inst2]{organization={Istituto di Scienza e Tecnologie dell'Informazione ``A. Faedo'', Consiglio Nazionale delle Ricerche (ISTI-CNR)},
            addressline={Via G. Moruzzi 1}, 
            city={Pisa},
            postcode={56126}, 
            country={Italy}}
\author[inst3]{Liping Zhao}
\affiliation[inst3]{organization={Department of Computer Science},
            addressline={University of Manchester}, 
            city={Manchester},
            postcode={M13 9PL}, 
            country={UK}}
\cortext[cor1]{Corresponding authors: Liping Zhao at liping.zhao@manchester.ac.uk and Alessio Ferrari at alessio.ferrari@isti.cnr.it}

\begin{abstract}

\textbf{Context:} Requirements engineering (RE) researchers have been experimenting 
 with machine learning (ML) and deep learning (DL) approaches for a range of RE tasks, such as requirements classification, requirements tracing, ambiguity detection, and modelling. However, most of today's ML/DL approaches are based on \textit{supervised} learning techniques, meaning that they need to be trained using a large amount of task-specific labelled training data. This constraint poses an enormous challenge to RE researchers, as the lack of labelled data makes it difficult for them to fully exploit the benefit of advanced ML/DL technologies. 
\textbf{Objective:} This paper addresses this problem by showing how a \textit{zero-shot learning} (ZSL) approach can be used for requirements classification without using any labelled training data. We focus on the classification task because many RE tasks can be framed as classification problems.
\textbf{Method:} The ZSL approach used in our study employs contextual word-embeddings and transformer-based  language models (LMs). We demonstrate this approach through a series of experiments to perform three classification tasks: (1) FR/NFR --- classification functional requirements vs non-functional requirements; (2) NFR --- identification of NFR classes; (3) Security --- classification of security vs non-security requirements.
\textbf{Results:} The study shows that the ZSL approach achieves an F1 score of 0.66 for the FR/NFR task. For the NFR task, the approach yields F1$\sim 0.72 - 0.80$, considering the most frequent classes. For the Security task, F1 $\sim 0.66$. All of the aforementioned F1 scores are achieved with zero-training efforts. 
\textbf{Conclusion:} This study demonstrates the potential of ZSL for requirements classification. An important implication is that it is possible to have very little or no training data to perform classification tasks. The proposed approach thus contributes to the solution of the long-standing problem of data shortage in RE. 

\end{abstract}

\begin{highlights}

\item  Lack of large, labelled datasets is a main challenge in using any form of machine learning for automatic requirements classification.
\item  We propose to use Zero-shot learning (ZSL) for requirements classification, which requires no labelled data and no training.
\item  Our results show that ZSL achieves acceptable performance  (F1 from 66\% to 80\%) for binary and multi-class classification tasks.
\item 
ZSL with generic language models (LMs), and simple label configurations appear to outperform domain-specific LMs and complex label configurations

\end{highlights}


\begin{keyword}
Zero-shot learning \sep language models \sep contextual word-embeddings \sep requirements classification \sep zero-shot text classification \sep unsupervised learning \sep multi-label classification \sep transfer learning \sep deep learning \sep requirements engineering \sep AI for requirements engineering \sep AI for software engineering
\PACS 0000 \sep 1111
\MSC 0000 \sep 1111
\end{keyword}
\end{frontmatter}

\section{Introduction}
\label{sec:intro}

In requirements engineering (RE), system and software requirements specifications are typically written in natural language (NL) \cite{zhao2021natural,kassab2014state}. In the last few years, natural language processing (NLP) techniques based on supervised machine learning (ML), and, more recently, deep learning (DL), have been applied to address several RE tasks, driven by the success of these techniques in a range of domains, including medical diagnosis, credit card fraud detection, and sentiment analysis \cite{sidey2019machine,sarker2021machine,minaee2021deep}. 

To date, research on ML-based RE has been primarily focused on \textit{supervised} classification approaches \cite{binkhonain2019review}, as most RE tasks can be framed as \textit{text classification} problems solved by supervised learning techniques. 
Relevant examples are: classifying requirements into different categories \cite{cleland2007automated,kurtanovic2017automatically,hey2020norbert}; identifying requirements from software contracts \cite{sainani2020extracting}; discerning requirements and non-requirements \cite{abualhaija2020automated}; identifying miscategorised requirements~\cite{ferrari2017natural}; and discovering requirements-relevant content from app reviews \cite{dkabrowski2022analysing,maalej2016automatic}. 


However, supervised ML has some major limitations. The most notable one is that supervised learners need to be trained on a large amount of task-specific, labelled data before they can be ready for predicting the outcomes on new data \cite{sarker2021machine,wang2019survey}. This problem is exacerbated in domains like RE where collecting and labelling sufficient training data is often expensive, time-consuming and error-prone~\cite{dalpiaz2019requirements,sainani2020extracting}. Labelling data also requires substantial domain- and even project-specific knowledge~\cite{ferrari2017natural}. Furthermore, labelled data used in previous studies are frequently unavailable. This happens even for the lively area of app review analysis, in which most studies have not released their labelled dataset, according to a recent survey (cf. ~\cite{dkabrowski2022analysing}, p. 34). This limitation hinders RE researchers from exploring different learning techniques.

Another limitation of supervised learning methods is that models can only classify the data belonging to \textit{seen classes} (i.e., classes labelled in the training data), but they cannot classify the data into previously \textit{unseen classes} (i.e., classes not labelled in the training data) \cite{wang2019survey}. Although this limitation is inherent in supervised learning, the ability to deal with previously unseen classes can bring huge 
 benefits to many real-world applications where classes are artificially defined, with no common consensus, or where classes may evolve over time, with new classes emerging and old ones becoming obsolete. One such example is requirements classification where several classification schemes exist for non-functional requirements (NFRs) \cite{glinz2007non,eckhardt2016non}. 

As software applications, their requirements, and the theory of NFRs itself evolves over time, so do the classification schemes. Consequently, a dataset labelled using one set of classes (e.g., the PROMISE NFR Dataset \cite{jane_cleland_huang_2007_268542}) cannot be reused to train a method that intends to predict a different set of classes (e.g., based on the latest ISO/IEC/IEEE 29148 standard \cite{ieeestandard}). Each time a new classification scheme is used, the datasets must be relabelled accordingly, incurring expensive data-labelling costs.

To address these problems, different learning paradigms have been proposed in recent years \cite{sarker2021machine}. One such paradigm is \textit{transfer learning} \cite{pan2009survey},  which aims to alleviate the problems of data shortages and expensive data labelling efforts by adapting existing well-trained ML models to different, but related domains or tasks \citep{zhuang2020comprehensive}. However, model adaptation still requires thousands or tens of thousands of labelled task-specific instances \cite{devlin2018bert}.

More recently, \textit{zero-shot learning} (ZSL) has emerged as a promising paradigm \cite{wang2019survey}. ZSL directly applies previously trained models to predicting both seen and unseen classes without using any labelled training instances \cite{larochelle2008zero,lampert2009learning}.

Expanding on our preliminary study \cite{alhoshan2022zero}, this paper aims to conduct an in-depth study of using ZSL for requirements classification and to gain insight into this new paradigm in the context of RE. Whereas our preliminary study only assessed ZSL on the classification of security and usability requirements selected from a portion of the PROMISE NFR dataset \cite{jane_cleland_huang_2007_268542}, this paper substantially extends the previous contribution by evaluating ZSL on different classification tasks, namely differentiation between functional requirements (FRs) and non-functional requirements (NFRs), identification of different NFR classes, and classification of security vs non-security requirements. These tasks are carried out on two datasets: the full PROMISE NFR dataset and the SecReq dataset~\cite{knausseric20214530183}. In addition, we have selected four different language models (LMs) to evaluate ZSL, to allow us to compare the performance of ZSL under different models.

The remaining paper is structured as follows: Section \ref{sec:related} briefly reviews the current ML approaches for requirements classification. Section \ref{sec:ZSL} introduces  the zero-shot learning approach used in this paper and its related concepts. Section \ref{sec:exp} defines the research questions for our study and details our study design. Section \ref{sec:res} analyzes the experimental results, while Section \ref{sec:find} answers our research questions based on these results. Section \ref{sec:threats} examines the validity threats to our experiments and our mitigation strategies. Finally, Section \ref{sec:con} concludes the paper.

\section{Related Work}
\label{sec:related}

Most studies in ML-based requirements classification focus on the categorisation between functional (FR) and non-functional (NFR, or ``quality''~\cite{li2014non}) requirements, and on the further categorisation of different NFR classes, such as security, performance, usability, \textit{etc.} However, the distinction between FR and NFR has been a matter of debate in the RE community~\cite{broy2015rethinking,dalpiaz2019requirements}, and the empirical study by Eckhardt~\textit{et al.}~\cite{eckhardt2016non} shows that NFRs can include functional aspects. Furthermore, there is a more fine-grained representation of FRs and NFRs given by the ISO/IEC/IEEE 29148:2018(E) Standard~\cite{ieeestandard}, which distinguishes between functional/performance, quality, usability, interface, and other classes, thus refining the conceptualisation already elaborated by the NFR classification from Glinz~\cite{glinz2007non}. Yet, despite the lack of consensus what NFRs are, and how we should classify and represent them, the differentiation between FRs and NFRs is a common categorisation in RE, and in the following we will use this distinction, keeping in mind that it is an artificial construct~\cite{eckhardt2016non}.

ML-based approaches for requirements classification were examined in a systematic literature review by Binkhonain and Zhao \cite{binkhonain2019review}. Here we briefly review some closely related representative works.

\subsection{Classification of FRs and NFRs}

One of the earliest adoptions of ML to RE was due to Cleland-Huang \textit{et al.}~\cite{cleland2007automated}, who proposed to use a set of indicator terms to identify different classes of NFR. The approach was supervised, in that it first identified a set of indicator terms on a set of manually annotated requirements, and then used this set to classify unseen cases. The approach achieved a recall up to 0.80, but suffered from low precision, up to 0.21. This study also introduced the PROMISE NFR dataset~\cite{jane_cleland_huang_2007_268542}, which has been widely used by the research community, and it is also one of the benchmarks of our work. 

To mitigate the problem of dataset annotation, Casamayor \textit{et al.}~\cite{casamayor2010identification} proposed a semi-supervised method, based on an iterative process similar to active learning, in which the user provided feedback to the classifier. Their approach used Naive Bayes (NB) as a classification algorithm and the PROMISE NFR dataset as the training set. After multiple iterations in which an increasing number of training examples were used, they obtained a maximum precision of above 0.80 and a maximum recall of above 0.70 on most classes, except underrepresented ones.

Another well-known ML approach is provided by Kurtanovi\'c and Maalej~\cite{kurtanovic2017automatically}, who applied Support Vector Machine (SVM) for requirements classification. They selected relevant features with an ensemble of different supervised classifiers and achieved precision and recall up to 0.92 for identifying FRs and NFRs  on the PROMISE NFR dataset. For the identification of specific NFRs classes, they achieved the highest precision and recall for security and performance classes with 0.92 precision and 0.90 recall. Dalpiaz \textit{et al.}~\cite{dalpiaz2019requirements} reconstructed the study by Kurtanovi\'c and Maalej and used the results obtained as a baseline to evaluate their proposed approach using interpretable linguistic features.

To overcome the problem of labour intensive feature engineering, Navarro \textit{et al.}~\cite{navarro2017towards} proposed one of the first approaches using a deep learning (DL) model. 
They used a CNN (Convolutional Neural Network) model on the PROMISE dataset, and obtained precision and recall of 0.80 and 0.79 respectively, thus addressing the problem of limited precision observed by Cleland-Huang \textit{et al.}. Similar approaches were proposed by Dekhtyar and Fong~\cite{dekhtyar2017re}, and more recently, by Aldhafer \textit{et al.}~\cite{aldhafer2022end}.


A more closely related work to ours is Hey \textit{et al.}~\cite{hey2020norbert}, who proposed NoRBERT, a transfer learning approach for requirements classification. Their approach is based on fine-tuning the BERT model (Bidirectional Encoder Representations from Transformers) \cite{devlin2018bert}. They achieved similar or better results with respect to previous works, achieving 0.92 precision and 0.95 recall for FR vs NFR classification on the PROMISE dataset.
NoRBERT also outperformed recent approaches at classifying NFRs classes. The most frequent classes were classified with  precision up to 0.94 and recall up to 0.90. 
The proposed solution was also applied for the classification of different types of functional requirements concerns in PROMISE, achieving  precision up to 0.88 and recall up to 0.95.



\subsection{Classification of Security Requirements}

One of the early works on security requirements classification was by Knauss \textit{et al.}~\cite{knauss2011supporting}, who used a Bayesian classifier to identify security-relevant requirements on three industrial datasets. These datasets are also used in our paper (aggregated into the \textit{SeqReq} dataset). They achieved precision $> 0.8$ and recall $>0.9$. In another work, Riaz \textit{et al.}~\cite{riaz2014hidden} proposed an approach to extract security-relevant sentences from requirements documents. They used a dataset of 10,963 sentences belonging to six different documents from the healthcare domain. The proposed approach was semi-automatic and based on KNN (K-nearest Neighbours) classification. The authors achieved a precision of $0.82$, and a recall of $0.79$.

Addressing the lack of domain-specific data sets, Munaiah \textit{et al.}~\cite{munaiah2017domain} proposed a domain-independent classification model for identifying domain-specific security requirements. The proposed approach, a one-class SVM classifier, was used to identify general descriptions related to software security weaknesses, but not the actual security requirements \textit{per se}, as the classifier was trained using the Common Weakness Enumeration database \citep{christey2013common}. 
The authors showed that the one-class classifier achieved an average precision and recall of 0.67 and 0.70 respectively.
Varenov \textit{et al.}~\cite{varenov2021security} compared the performance of different LMs, namely BERT, XLNET, and DistilBERT, for security requirements classification. They identified $1,086$ security requirements of seven different classes collected from multiple existing datasets, such as PURE~\cite{ferrari2017pure}, SecReq~\cite{knauss2011supporting} and Riaz's dataset~\cite{riaz2014hidden}. Unlike  previous studies, the work by Varenov \textit{et al.}~\cite{varenov2021security} aimed to classify security requirements into more fine-grained classes, i.e., Confidentiality, Integrity, Availability, Accountability, Operational, Access Control, and Other. DistilBERT achieved the best results, with precision of 0.80 and recall of 0.82.

\subsection{Our Contribution}
In comparison with the  related work, our study aims to present a comparative analysis of different ZSL configurations for the classification of requirements. Similarly to the proposal of Hey \textit{et al.}~\cite{hey2020norbert}, we explore the potential of a DL solution on the widely used PROMISE dataset. 
Differently from Hey \textit{et al.}~\cite{hey2020norbert}, this is the first work in RE that proposes to use ZSL for the classification task. While Hey \textit{et al.} focus on addressing generalisability of the classifier by means of transfer learning, our proposal: (1) avoids the need of a tagged dataset, therefore addressing the well known problem of the scarcity of annotated datasets in RE~\cite{zhao2021natural,ferrari2017natural,dalpiaz2018natural,ferrari2017pure}; (2) is inherently generalisable to different projects, thus addressing the problem of decreasing performance with unseen projects, which typically affects requirements classifiers~\cite{hey2020norbert,dalpiaz2019requirements}. Concerning security requirements classification, our proposal overcomes the problem of dataset annotation as Munaiah \textit{et al.}~\cite{munaiah2017domain}. However, their approach is specific to security requirements, while our proposal is more generalisable and adaptable to different classification tasks.

\section{Zero-Shot Learning}
\label{sec:ZSL}

Zero-short learning is an emerging learning paradigm that aims to perform learning tasks without using training data. ZSL was originally used in image processing to predict unseen images \cite{romera2015embarrassingly}, but has recently been adapted to many NLP tasks, including entity recognition \cite{ma2016label}, relation extraction \cite{levy2017zero}, document classification \cite{nam2016all}, and text classification \cite{pushp2017train}. The fundamental idea of ZSL is that some previously trained language models are so accurate that they can be directly applied to predicting new data without any training \cite{larochelle2008zero,lampert2009learning}. In this section, we first introduce the concepts of language models and then focus on a specific ZSL approach---embedding-based ZSL---used in our study.

\subsection{Language Models and Transfer Learning}
\label{sec:LM}
 
Language models (LMs) are deep neural networks for representing words and sentences in natural language. These models are developed to support NLP tasks such as language understanding and inference. Traditional LMs, such as Skip-gram \cite{mikolov2013distributed}, Word2Vec \cite{mikolov2013efficient}, and GloVe \cite{pennington2014glove}, use \textit{static word embeddings}, i.e., fixed vector representations, to represent the words and sentences in a text \cite{mikolov2013distributed}. More recently, \textit{transformer-based LMs}, such as OpenAI GPT \cite{radford2018improving} and BERT \cite{devlin2018bert}, have made significant improvements over the traditional LMs in language representation, as they can capture the deep meaning of words and sentences through \textit{dynamic or contextual word embeddings} \cite{ethayarajh2019contextual}. In other words, traditional LMs are \textit{static} as they simply expand the words in a sentence with related ones. For example, the sentence ``\textit{This is not about usability}'' is mapped onto a vector similar to ``\textit{This is about usability}'', as the two sentences only differ in one word. By contrast, with transformer-based LMs, the vector representations of these two sentences are different, as they have opposite meanings. This characteristic naturally plays a crucial role in requirements analysis, as requirements sentences often use a very restricted vocabulary~\cite{ferrari2017pure}, but convey different meanings. 

Today, pretrained LMs (PLMs) are widely available\footnote{For example, a large number of PLMs are freely available at Hugging Face website: \url{https://huggingface.co/models}.}. These PLMs are typically pretrained for some generic NLP tasks using unlabeled data such as app reviews and Wikipedia, available \textit{en masse} through the Internet \cite{raffel2020exploring}. Such LMs can be adapted--- i.e., \textit{transferred}---to different downstream tasks \cite{radford2019language}. This concept is known as \textit{transfer learning}~\cite{ruder2019transfer}.

 There are two approaches for adapting a PLM: \textit{feature-based} and \textit{fine-tuning} \citep{ruder2019neural}. The feature-based approach uses the representations of a pre-trained model as input features to train a downstream task model (e.g., ELMo \cite{peters-etal-2018-deep}), while the fine-tuning approach, adopted by OpenAI GPT and BERT, involves modifying the weights and parameters in certain layers of the PLM so as to enable the model to perform a specific NLP downstream task. Transfer learning helps to reduce the cost and effort required for training new models, and allows users to explore different NLP tasks with a relatively small amount of task-specific labelled data.

\subsection{Embedding-Based Zero-Shot Learning}

There are two common approaches to ZSL: \textit{entailment-based} and \textit{embedding-based}. The former treats the classification task as a natural language inference (NLI) task \cite{yinroth2019zeroshot}, whereas the latter uses language representations to predict if an input text is related to a given class label. More specifically, the entailment-based approach treats an input text sequence as a premise and the labels as a hypothesis, and then infers if the input text is an entailment of any of the labels or not \cite{yinroth2019zeroshot}. For example, given the sentence \textit{``the system must be deployed on Azure''} as a premise, and the label string \textit{``this is about software architecture''} as a hypothesis, the entailment-based classifier provides a score which is then translated into one of the following outputs: entailment (yes), contradiction (no), or undecided. This ZSL approach requires a large inference-based PLM\footnote{Hereafter we simply call a ``PLM'' ``LM'', as the LMs used in our study are PLMs.} that can interpret the entailment relation between an input sequence and a label.

The embedding-based ZSL approach was introduced by Veeranna \textit{et al.}~\cite{sappadla2016using}. Under this approach, both class labels (e.g., ``usability'', ``security'') and input text are represented as word sequences using word embeddings. Text classification then involves computing the \textit{semantic similarity} between each label sequence and the text sequence. If the similarity score is greater than a certain threshold, then the text can be classified into a specific category represented by the label; otherwise, the text does not belong to that category. Note that, since a label is treated as a sequence of words, it can contain any number of words or their combinations. 

Due to the simplicity of the embedding-based approach to ZSL, we have applied it in our study. However, unlike the original proposal by Veeranna \textit{et al.} \cite{sappadla2016using}, who used the static word-embedding technique (Skip-gram) for word representation, we take advantage of transformer-based LMs and use them to produce contextual word-embeddings for both labels and input text. In so doing, our embeddings not only capture syntactic and semantic characteristics of the words, but also their context. 

Another difference between our approach and the original one is that we do not use similarity thresholds to determine the predicted label; instead, we treat all text classification as a multi-label classification task and rank-order all the labels with their similarity scores. For a binary or multi-class classification task, we select the label with the highest similarity score as the predicted label. For a multi-label task, we check the top-$n$ labels.  The usage of similarity thresholds could help identify misclassifications, when the highest similarity has a low value. However, selecting appropriate thresholds is a task \textit{per se}, which we do not consider in this paper.

\begin{figure}[t!]
  \centering
    \includegraphics[width=\textwidth]{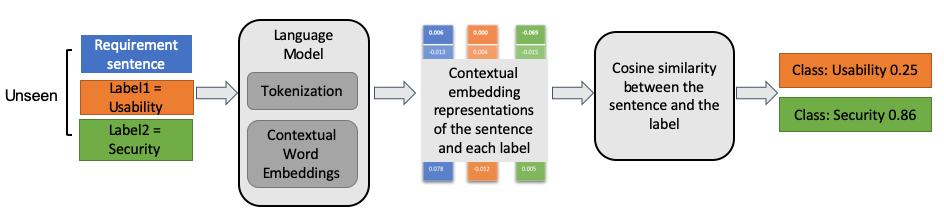}
    \caption{An Illustration of the Contextual Word Embedding-Based ZSL Approach.}
    \label{ZSLApproach}
\end{figure}

The \textit{contextual word embedding-based} approach adopted in our study (Figure \ref{ZSLApproach}) can be explained using a simple example: Given a requirement ``\textit{CNG shall support mechanisms for secure authentication and communication with the remote management system}'' and two class labels as ``usability'' and ``security'', we want to find out if the requirement should be classified into the ``usability'' or the ``security'' class. ZSL performs this classification task by taking the three word sequences (the requirement statement plus the two labels) as input to a PLM to produce three contextual word embeddings. It then compares the requirement with each label using the $Cosine$ similarity function\footnote{The $Cosine$ similarity function is the standard way to compute semantic similarity between a label and a text \cite{mikolov2013efficient,pennington2014glove,bojanowski2017enriching,ethayarajh2019contextual}.}. The comparison will return $n$ similarity scores, each score for a label and the requirement pair. The pair with the highest similarity score means that the requirement belongs to the category denoted by its associated label. In our example, ZSL would return
two similarity scores: $0.86$ for the ``security'' label and the given requirement, and $0.25$ for the ``usability'' label and the given requirement. Based on these scores, we deduce that the given requirement is a security requirement.

The above example shows that the accuracy of the embedding-based ZSL approach depends highly on the choice of 1) the \textbf{labels} and 2) the \textbf{PLMs}. Whereas the above example only uses single word labels (i.e., ``usability'' and ``security''), our study will investigate different label configurations. For example, by composing the usability label using a set of synonyms and related words, such as instructive, easy, helpful, useful, learnable, explainable, intuitive, and understandable, the LM can produce a more dynamic embedding that can capture a range of connotations of the usability requirements.

\vspace{1em}

\section{Experimental Design}
\label{sec:exp}

To evaluate the effectiveness of embedding-based ZSL for requirements classification, our study aims to answer the following three research questions (RQs):

\begin{description}
    \item[\textbf{RQ1:}]  \textit{Which \textbf{language model} is more effective for which zero-shot requirements classification task?}

    \item[\textbf{RQ2:}] \textit{To what extent do different \textbf{label configurations} affect the effectiveness of zero-shot requirements classification?}

    \item[\textbf{RQ3:}] \textit{How \textbf{effective} is zero-shot learning for requirements classification  compared to related supervised learning approaches?}
\end{description}

 We answer these questions through a series of experiments. In this section, we describe our experimental design, consisting of five steps: selection of datasets and tasks; selection of LMs; label configuration; performance measure selection; technical setup of the experiments. 

\subsection{Dataset and Task Selection}

The following two datasets are selected for our experiments: 

\begin{itemize}
    
    \item \textbf{PROMISE NFR dataset} \cite{jane_cleland_huang_2007_268542}, introduced by Cleland Huang \textit{et al.} \cite{cleland2007automated}: This dataset contains 625 requirements, partitioned into 255 FRs, and 370 NFRs. The NFRs are further partitioned into 11 different classes, namely: A = Availability (21 requirements), L = Legal (13), LF = Look and feel (38), MN = Maintainability (17), O = Operational (62), PE = Performance (54), SC = Scalability (21), SE = Security (66), US = Usability (67), FT = Fault tolerance (10), and PO = Portability (1). These classes are unevenly distributed, ranging from 67 requirements for Usability to one for Portability. Each one of the most frequent classes---Usability, Security, Operational, and Performance---has more than 50 examples, while the less frequent classes---Fault Tolerance, Legal, Maintainability and Portability---have from one to 17 requirements each. The dataset has been widely used in the literature, e.g., by Kurtanovi{\'c} and Maalej \cite{kurtanovic2017automatically}, and by Hey \textit{et al.} \cite{hey2020norbert}.
    
    \item \textbf{SecReq dataset} \cite{knausseric20214530183}, introduced by Knauss \textit{et al.} \cite{knauss2011supporting}: This dataset contains 510 requirements, made of security-related requirements (187) and non-security related requirements (323). The requirements were collected from three projects: Common Electronic Purse (ePurse), Customer Premises Network (CPN), and Global Platform Spec (GPS). The dataset has been used, e.g., by Varenov \textit{et al.}~\cite{varenov2021security}.

\end{itemize}

We select the following typical requirements classification tasks for our study:

\begin{itemize}
    \item \textbf{Task FR/NFR}---\textit{Binary Classification of FRs vs. NFRs.}  
    With this task we aim to distinguish FRs from NFRs, assuming that a requirement belongs to either a FR or a NFR class. We use the PROMISE NFR dataset for this task.
    
    \item \textbf{Task NFR}---\textit{Binary, Multi-class and Multi-label Classification of NFRs}. This task aims to classify different types of NFRs based on the 10 different classes of the PROMISE NFR dataset (we excluded the Portability class as it only has one single sample in the dataset). We perform three sub-tasks to understand how ZSL reacts to different ways of classifying NFRs: 1) binary classification which discerns if a NFR belongs to a particular class or not; 2) multi-class single-label classification (simply, \textit{multi-class classification}) which assigns a NFR to one of the top  or all NFR classes; 3) multi-class multi-label classification (simply, \textit{multi-label classification}), which allocates a NFR to one or more NFR classes. The purpose of the third sub-task is to check if the top-\textit{n} NFR classes returned by the ZSL classifier correlate with the assigned NFR label in the datatset.  
    
    \item \textbf{Task Security} ---\textit{Binary Classification of security related vs. non-security related requirements.} This task assumes that a requirement belongs only to one of these two classes: security related and non-security related. We use the SeqReq dataset for this task.
\end{itemize}

These datasets and tasks are selected for our experiments as they are frequently considered in the literature (cf. Sect.~\ref{sec:related}) and will enable us to compare our results directly with those obtained by previous work. 

\subsection{Language Model Selection}

 We select the following four BERT-based LMs for our study: two generic and two domain-specific LMs. The two generic LMs are \textbf{Sentence-BERT (Sbert)} and \textbf{All-MiniLM-L12 (AllMini)}, which are freely avaiable at the HuggingFace website\footnote{\url{https://huggingface.co}}, a well-known NLP community repository that provides open source pretrained LMs and other language resources. The two domain-specific LMs, \textbf{Bert4RE} \cite{ajagbe2022RE} and \textbf{BERTOverflow (SObert)} \cite{tabassum-etal-2020-code}, were developed for requirements and software engineering tasks.
 
 We focus on the BERT-based models, due to their popularity and suitability for requirements classification~\cite{hey2020norbert}. Other LMs, such as GPT-2 and GPT-3 by OpenAI, and XLNet \cite{yang2019xlnet}, have not been included in our study, as they are not suitable for requirements classification.  
 For example, GPT-2 and GPT-3 are mainly for language generation tasks, such as language translation and text summarization~\cite{brown2020language}, whereas XLN-Net is for NLP involving processing long texts such as paragraphs~\cite{yang2019xlnet}. Requirements classification typically deals with texts at the sentence level. Below we introduce the four LMs used in our experiments.

    \begin{itemize}

        \item \textbf{Sbert:} This generic LM\footnote{\url{huggingface.co/deepset/sentence\_bert}}, proposed by Reimers and Gure\-vych~\cite{reimers2019sentence}, is a fine-tuned version of BERT LM which aims to enrich the semantic embedding representation, i.e., to aid in deriving semantically meaningful sentence embeddings. 
        The LM overcomes the drawbacks of the original BERT models, which use word embeddings to generate sentence embeddings and thus result in weak semantic representations of sentences~\cite{reimers2019sentence}.  
       
        \item \textbf{AllMini:} Introduced by Wang \textit{et al.}~\cite{wang2020minilm} at Microsoft Research, this LM\footnote{\url{huggingface.co/sentence-transformers/all-MiniLM-L12-v2}} aims to overcome the complexity of some LMs such as BERT models which usually consist of millions of parameters and can be challenging for pre-training and fine-tuning. AllMini reduces (or \textit{distils}) the size of the BERT models, while preserving their performance. The main purpose of AllMini is to support sentence embeddings. In this experiment, we use a version of AllMini (\textbf{All-MiniLM v2)}, which was fine-tuned using one billion sentence pairs. This LM is used for encoding sentences and short paragraphs and is particularly efficient for semantic search and sentence clustering tasks. 
        
        \item \textbf{Bert4RE:}	This is a RE domain-specific LM \cite{ajagbe2022RE} which was trained on the BERT\textsubscript{base} model using more than seven million words from different RE-related datasets, including the PROMISE NFR dataset, the PURE dataset \cite{ferrari2017pure}, and app reviews from Google Playstore and App Store. Although Bert4RE aims to support a wide range of RE tasks, it has only been tested on the task of identifying semantic roles from requirements documents. As this is the only publicly available RE-specific LM, we include it in our study. The BERT4RE LM is provided by the authors in a Zenodo repository \footnote{\url{zenodo.org/record/6354280}}. 
        
        \item \textbf{SObert:} This is a SE domain-specific LM \cite{tabassum-etal-2020-code} that was trained on 152 million sentences from \textsc{Stack Overflow}\footnote{\url{https://stackoverflow.com/}}. SObert\footnote{\url{huggingface.co/jeniya/BERTOverflow}} shares the same vision as Bert4RE, aiming to capture semantics of the SE terminology. Although SObert has been trained to perform SE specific \textit{named entity recognition (NER)} tasks, it is among the few SE specific LMs that can potentially be adopted for requirements classification\footnote{Note that another LM for SE is CodeBERT~\cite{feng-etal-2020-codebert}, but it was trained on both natural language and programming texts for specific tasks that involve code retrieval based on natural language queries, which are not suitable for requirements classification.}.
    \end{itemize}

\subsection{Label Creation and Configuration}
\label{sec:labels}

Different label creation strategies are used to produce the labels for each requirements class, described as follows. 

\begin{itemize}

    \item \textbf{Original labels:} These labels were \textit{derived} from the original class names used in the dataset without using any external knowledge. For example for the task FR/NFR, the original label for the class FR is ``functional'', while for NFR we use two types of label: 1) the expression ``not about functional''\footnote{Preliminary experiments have shown that this term is more effective for ZSL with respect to ``non functional''.} and 2) a string including all the NFR class names (``usability, security, availability, ...''). 
    
    \item \textbf{Expert curated labels:} These labels were \textit{curated} by the three authors of this paper based on their understanding of the requirements classes. The curation process took three steps: First, we independently provided a set of terms to describe each requirements class, resulting in three sets of terms per class. Second, we discussed our selections and produced a set for each class together. Third, we performed preliminary trials on a part of the requirements, to select the best subset of expert-curated labels. 
    For example, for the task FR/NFR, the label for the class FR is composed of the terms ``functional, system, behavior, shall, must'', which are typically associated with FRs.  
    These expert-curated labels are expected to complement the aforementioned original labels as they can better discriminate requirements classes. 

    \item \textbf{Word-embedding generated labels:} These labels consisted of the terms \textit{extracted} and \textit{selected} from word-embeddings learned from the text of Wikipedia pages belonging to the Computer Science (CS) portal. The idea was that the embeddings learned from the CS portal represent the meaning of words in the CS domain, and are therefore more suitable than a generic LM in providing similar terms for our CS context. We adopted the embedding approach and code provided by Ferrari and Esuli~\cite{ferrari2019nlp}. To agree on a label for each requirements class, we followed these steps: (1) The top-\textit{n} most similar terms were selected according to the word-embeddings; (2) each of the three authors independently annotated the terms with \textit{yes} (indicating the term to be representative for the class), \textit{no} (indicating the term not to be representative for the class), or \textit{maybe} (indicating that the term could be representative for the class); (3) each author revised their ``maybe'' answer as either \textit{yes} or \textit{no}. 
    (4) the final answer for each class was decided through majority voting. 
    After this procedure, the terms that were tagged with \textit{yes} were included in the label. 
    To assess the degree of agreement between the three authors in step (2), we evaluate the overall percentages of agreement on each term (i.e., all annotators tagged the term with \textit{yes} or \textit{no}), partial agreement (i.e., two out of the three annotators tagged the term with \textit{yes} or \textit{no}), and disagreement (i.e., the three annotators selected three different tags: \textit{yes}, \textit{no}, and \textit{maybe}). In step (3), after resolving the tags \textit{``maybe''}, we consider the inter-rater agreement (IRR) based on Krippendorff’s alpha test \cite{krippendorff2018content} as well as Fleiss’s Kappa \cite{fleiss1973equivalence}. These statistical tests are used to measure the level of agreement among us. The interpretation of the test results follows the guidelines reported in the Koch Kappa benchmark \cite{landis1977measurement}. 
\end{itemize}    

The above three label creation strategies are combined into different configurations which are then adopted to each specific classification task to produce task specific labels. Different label configurations and their associated classification tasks are reported in Section~\ref{sec:res} for each task. 

\subsection{Performance Measures}

For each LM and its label configuration, we measure their performance on each class with respect to a specific classification task using both unweighted and weighted precision (P), recall (R), and F1-score (F1). The weighted P, R and F1 (represented respectively as wP, wR and wF1) are calculated using the distribution of the NFR classes in the dataset. Basically, we performed a weighted sum of the different measures, where the weights are the percentages of requirements in a certain class. These weighted results enable us to compare our results with other studies in requirements classification~\cite{hey2020norbert,kurtanovic2017automatically}. Instead, the unweighted values, reported by each class, allow us to provide more-fine grained analyses, especially for the multi-class and multi-label cases.

\subsection{Experimental Setup}

Based on the combinations of different LMs, different label selection strategies, and different classification tasks, we have designed and conducted more than 360 experiments. We set up each experiment as a combination of one of the selected LMs, one specific label configuration and one specific task for each dataset. We call each LM-Label combination as a \textbf{ZSL classifier}. 
We use the Transformer API Python package\footnote{\url{huggingface.co/docs/transformers}} to import and prepare the four selected LMs with their transformer-like tokenizers, and we use the $torch.nn$ module\footnote{\url{pytorch.org/docs/stable/generated/torch.nn.CosineSimilarity.html}} in PyTorch to compute the cosine similarity score between two tensors (i.e., the PyTorch tensor objects obtained from the contextual representation by the selected LMs). The \textit{sklearn.metrics} module \cite{scikit-learn} is used to calculate the classification performance results in terms of the P, R and F1 scores.

\section{Experimental Results}
\label{sec:res}

In this section, we first report our label configurations and the IRR scores achieved from our label selection for the word-embedding strategy; we then report the experimental results obtained from using the best label configurations. 

\subsection{Task FR/NFR}
For the FR/NFR task we perform a binary classification, which aims to classify a requirement as either FR or NFR.


\subsubsection{Label Configuration}

\begin{table}
\centering
\caption{Label configurations for \textit{Task FR/NFR}. }
\label{tab:labelsFRvsNFR}
\resizebox{1\textwidth}{!}{
\begin{tabular}{llll} 
\toprule
\textbf{Label Abbr.}    & \textbf{Label Configuration}                                                                                         & \textbf{FR Label }                                                                                                               & \textbf{NFR Label }                                                                                                                                                                            \\ 
\hline
\textbf{\textit{FR\_A}} & Original 1                                                                                                           & ``functional''                                                                                                                     & ``not about functional''                                                                                                                                                                         \\ 
\hline
\textbf{\textit{FR\_B}} & Expert curated                                                                                                       & \begin{tabular}[c]{@{}l@{}}``functional, system, behavior, \\shall, or must''\end{tabular}                                         & \begin{tabular}[c]{@{}l@{}} ``not about functional, system, behavior,\\~shall, or must''\end{tabular}                                                                                             \\ 
\hline
\textbf{\textit{FR\_C}} & \begin{tabular}[c]{@{}l@{}}Word embedding \\(selected from top 20 words) +\\~Expert curated\end{tabular}             & \begin{tabular}[c]{@{}l@{}}``functional, system, behavior, \\shall, must, procedural, structural, \\or characterize''\end{tabular} & \begin{tabular}[c]{@{}l@{}}``not about functional, system, behavior, \\shall, must, procedural, structural, \\or characterize''\end{tabular}                                                     \\ 
\hline
\textbf{\textit{FR\_D}} & Original 2                                                                                                           & ``functional''                                                                                                                     & \begin{tabular}[c]{@{}l@{}}``usability, security, availability, legal, look \& feel,\\scalability, fault tolerance, performance,\\operational, maintainability, or portability''\end{tabular}    \\ 
\hline
\textbf{\textit{FR\_E}} & \begin{tabular}[c]{@{}l@{}}Expert curated +\\~Original 2\end{tabular}                                                & \begin{tabular}[c]{@{}l@{}}``functional, system, behavior,\\~shall, or must''\end{tabular}                                         & \begin{tabular}[c]{@{}l@{}}``usability, security, availability, legal, look \& feel,\\scalability, fault tolerance, performance,\\operational, maintainability, or portability''\end{tabular}    \\ 
\hline
\textbf{\textit{FR\_F}} & \begin{tabular}[c]{@{}l@{}}Word embedding \\(selected from top 20 words) +\\Expert curated + Original 2\end{tabular} & \begin{tabular}[c]{@{}l@{}}``functional, system, behavior, \\shall, must, procedural, structural, \\or characterize''\end{tabular} & \begin{tabular}[c]{@{}l@{}}``usability, security, availability, legal, look \& feel, \\scalability, fault tolerance, performance, \\operational, maintainability, or portability''\end{tabular}  \\
\bottomrule
\end{tabular}
}
\end{table}

The label configurations for the FR/NFR Task are reported in Table~\ref{tab:labelsFRvsNFR}. The labels consist of two groups, one group represents the FR class, and the other the NFR class. Six configurations are used, which combine the different strategies discussed in Sect.~\ref{sec:labels}. 
In the selection of the word-embedding generated terms, the annotation procedure produced the following statistics: 75\% perfect agreement, 25 \% partial agreement, and 0\% disagreement. 
We also computed the IRR, and we obtained 0.41 as Krippendorff's alpha and a Fleiss' kappa score of 0.40, indicating a moderate agreement. 

Concerning the other strategies, it is worth remarking the usage of ``functional'' vs ``not about functional'' (strategy, FR\_1 Original 1). This type of strategy, in which the original label is negated with the prefix ``not about'' is also applied for the NFR class label of FR\_B and FR\_C, and will be applied also later on in this paper to represent the negation of a class in a binary classification. 

\subsubsection{FR vs NFR Binary Classification}
Table~\ref{tab:resultsFRvsNFRdet}, column Total, reports the overall classification results for all LMs and labelling strategy combinations. In \textbf{bold}, we highlight the best combination for each LM. 

The overall best combination is Sbert + FR\_E, achieving a wF1 score of 0.66, with wP = 0.71 and wR = 0.66. This indicates that the domain agnostic Sbert model, designed to provide a semantic-laden representation for generic sentences, substantially outperforms the other models for this task. Furthermore, the best labelling strategy for Sbert is FR\_E, i.e., the one that uses the Expert curated labels + Original labels, which identifies the NFRs using the names of the NFR classes (Usability, Security, Availability, \textit{etc.}).

\begin{table}[t]
\centering
\scriptsize
\caption{Classification results for FR and NFR classes on \textit{Task FR/NFR}, obtained from the best combination for each LM and label configuration.}
\label{tab:resultsFRvsNFRdet}

\begin{tabular}{l|lll|lll|lll} 
\toprule
\multirow{2}{*}{\textbf{ZSL Classifier}} & \multicolumn{3}{l|}{\textbf{Total}} & \multicolumn{3}{l|}{\textbf{FR (255)}}  & \multicolumn{3}{l}{\textbf{\textbf{NFR (370)}}}\\ 
\cline{2-10}
& \textbf{wP} & \textbf{wR} & \textbf{wF1} & \textbf{P} & \textbf{R} & \textbf{F1} &\textbf{P} &\textbf{R} &\textbf{F1}     \\ 
\hline
\textbf{Sbert + FR\_E}                       & \textbf{0.71} & \textbf{0.66} & \textbf{0.66} & \textbf{0.55}  & \textbf{0.82} & \textbf{0.66} & \textbf{0.82}  & 0.54          & 0.65              \\
AllMini + FR\_D                                & 0.63 & 0.59 & 0.59 & 0.50           & 0.71          & 0.58          & 0.72           & 0.50          & 0.59             \\
Bert4RE + FR\_C                               & 0.58 & 0.56 & 0.57 & 0.47           & 0.65          & 0.54          & 0.67           & 0.50          & 0.57              \\
\textbf{SObert + FR\_C}                      & 0.58 & 0.59  & 0.58    & 0.50           & 0.41          & 0.45          & 0.64           & \textbf{0.72} & \textbf{0.68}  \\
\bottomrule
\end{tabular}

\end{table}

Looking at Table~\ref{tab:resultsFRvsNFRdet}, last six columns, we can see how the performance is divided between FR and NFR classification. We see that the model tends to have higher precision on NFR (P = 0.82), and higher recall on FR (R = 0.82). This is an interesting result, as FRs are less frequent in the dataset (255 FR, 370 NFR), and one would expect to have the opposite result. Indeed, the most frequent class is typically returned more frequently in ML approaches, as it happens, e.g., for NoRBERT (cf. Hey \textit{et al.}~\cite{hey2020norbert}, Table III of their paper). This phenomenon occurs also for the other best configurations of LMs. This highlights a  characterising element of ZSL: the performance does not depend on the size of the dataset for each class, because no actual \textit{learning} is performed on the tagged data.
A further increase in the accuracy on the FR class could be potentially achieved with a more project-specific labelling strategy for functional requirements (i.e., choosing terms that characterise functional requirements in the specific project).  

\subsection{Task NFR}

In this task, we performed three classification sub-tasks: a binary classification to detect a specific NFR category (e.g., ``usability'' vs ``other''); a multi-class classification to classify a requirement into one class out of a set of NFR classes; a multi-label classification, in which each requirement is associated with a ranked list of NFR classes, and we want to see if the correct label is in the top-\textit{k} classes. This last approach can be applied in a semi-automatic classification context, in which the two top-\textit{k} classes are shown to the requirements analysts, and they are asked to select the correct one. For all the sub-tasks, we evaluate the results: 1) considering only requirements in the largest classes, namely security, usability, performance and operational, which include the majority of the requirements; 2) considering all the classes, except the portability class, which includes one requirement only. For the multi-label classification case, we consider $k = 2$ when only the largest classes are considered, and $k = 3$ when all the classes are considered. 


\subsubsection{Label Configuration}
\label{sec:NFRlabels}

\begin{table}
\centering
\caption{Label configurations for Usability and Security classes for \textit{Task NFR}, binary classification case.}
\label{tab:labelsNFRbinary}
 \resizebox{1\textwidth}{!}{
\begin{tabular}{llll} 
\toprule
\vcell{\textit{\textbf{Label Abbr.}}} & \vcell{\textbf{Label Configuration}}                                                                                 & \vcell{\textbf{NFR Label}}                                                                                                                                                                                                                                                                                                                                                            & \vcell{\textbf{``Other'' Label}}                                                                                                                                                                                              \\[-\rowheight]
\printcelltop                          & \printcelltop                                                                                                         & \printcelltop                                                                                                                                                                                                                                                                                                                                                                         & \printcelltop                                                                                                                                                                                                               \\ 
\hline
\textbf{Usability} \\
\vcell{\textbf{\textit{US\_A}}}        & \vcell{Original 1}                                                                                       & \vcell{``usability''}                                                                                                                                                                                                                                                                                                                                                                    & \vcell{``not about usability''}                                                                                                                                                                                               \\[-\rowheight]
\printcelltop                          & \printcelltop                                                                                                         & \printcelltop                                                                                                                                                                                                                                                                                                                                                                         & \printcelltop                                                                                                                                                                                                               \\
\vcell{\textbf{\textit{US\_B}}}        & \vcell{Expert curated}                                                                                              & \vcell{\begin{tabular}[b]{@{}l@{}}``instructive, easy, helpful, useful, \\learnable, explainable, affordable,\\~intuitive, or understandable''\end{tabular}}                                                                                                                                                                                                                             & \vcell{\begin{tabular}[b]{@{}l@{}}``not about instructive, easy,\\helpful, useful, learnable, \\explainable, affordable, \\intuitive, or understandable''\end{tabular}}                                                       \\[-\rowheight]
\printcelltop                          & \printcelltop                                                                                                         & \printcelltop                                                                                                                                                                                                                                                                                                                                                                         & \printcelltop                                                                                                                                                                                                               \\
\vcell{\textbf{\textit{US\_C}}}        & \vcell{\begin{tabular}[b]{@{}l@{}}Word embedding\\~(selected from top 20 words )\end{tabular}}                             & \vcell{\begin{tabular}[b]{@{}l@{}}``accessibility, aesthetic, contextual,\\experience, satisfaction, HCI, UX, \\questionnaire, ease, or ergonomics''\end{tabular}}                                                                                                                                                                                                                       & \vcell{\begin{tabular}[b]{@{}l@{}}''not about accessibility, aesthetic,\\~contextual, experience, satisfaction,\\HCI, UX, questionnaire, \\ease, or ergonomics''\end{tabular}}                                                \\[-\rowheight]
\printcelltop                          & \printcelltop                                                                                                         & \printcelltop                                                                                                                                                                                                                                                                                                                                                                         & \printcelltop                                                                                                                                                                                                               \\
\vcell{\textbf{\textit{US\_D}}}        & \vcell{\begin{tabular}[b]{@{}l@{}}Word embedding \\(selected from top 20 words) +\\~Original 2\end{tabular}} & \vcell{\begin{tabular}[b]{@{}l@{}}``accessibility, aesthetic, contextual,\\experience, satisfaction, HCI, UX, \\questionnaire, ease, or ergonomics''\end{tabular}}                                                                                                                                                                                                                       & \vcell{\begin{tabular}[b]{@{}l@{}}``security, performance, operational, \\look feel, legal, fault tolerance, \\maintainability, scalability, availability,\\~or portability''\end{tabular}}                                    \\[-\rowheight]
\printcelltop                          & \printcelltop                                                                                                         & \printcelltop                                                                                                                                                                                                                                                                                                                                                                         & \printcelltop                                                                                                                                                                                                               \\
\vcell{\textbf{\textit{US\_E}}}        & \vcell{\begin{tabular}[b]{@{}l@{}}Word embedding \\(selected from top 50 words)\\~+ Original 2\end{tabular}}  & \vcell{\begin{tabular}[b]{@{}l@{}}``accessibility, aesthetic, contextual,\\experience, satisfaction, HCI, UX,\\questionnaire, ease, ergonomics, \\designer, evaluate, multimodal, \\practitioner, prototyping, preference,\\~personalization, suitability, focus, \\clarity, responsiveness, judgement,\\~feel, or helpful''\end{tabular}}                                               & \vcell{\begin{tabular}[b]{@{}l@{}}``security, performance, operational,\\~look feel,~~legal, fault tolerance,\\~maintainability, scalability, availability, \\or portability''\end{tabular}}                                   \\[-\rowheight]
\printcelltop                          & \printcelltop                                                                                                         & \printcelltop                                                                                                                                                                                                                                                                                                                                                                         & \printcelltop                                                                                                                                                                                                               \\ 
\hline
\textbf{Security} \\
\textbf{\textit{SE\_A}}                & Original 1                                                                                               & ''security''                                                                                                                                                                                                                                                                                                                                                                            & ``not about~security''                                                                                                                                                                                                        \\
\textbf{\textit{SE\_B}}                & Expert curated                                                                                                      & ``security, authorization, or protection''                                                                                                                                                                                                                                                                                                                                              & ``not about security, authorization, or protection''                                                                                                                                                                          \\
\textbf{\textit{SE\_C}}                & \begin{tabular}[c]{@{}l@{}}Word embedding~\\(selected from top 20 words)\end{tabular}                                     & \begin{tabular}[c]{@{}l@{}}``vulnerability, securing, protecting, \\protection, cybersecurity, assurance,\\~cyber, countermeasure, threat, privacy,\\~authentication, prevention, or confidentiality''\end{tabular}                                                                                                                                                                     & \begin{tabular}[c]{@{}l@{}}``not about vulnerability, securing, protecting,\\protection,~cybersecurity, assurance,\\~cyber, countermeasure, threat, privacy,\\~authentication, prevention, or confidentiality''\end{tabular}  \\
\textbf{\textit{SE\_D}}                & \begin{tabular}[c]{@{}l@{}}Word embedding\\(selected from top 20 words) +~\\Original 2\end{tabular}          & \begin{tabular}[c]{@{}l@{}}``vulnerability, securing, protecting,\\protection,~cybersecurity, assurance,~\\cyber, countermeasure, threat, privacy,~\\authentication, prevention, or confidentiality''\end{tabular}                                                                                                                                                                      & \begin{tabular}[c]{@{}l@{}}``usability, performance, operational, \\look \& feel, legal, fault \& tolerance,\\~maintainability, scalability, availability, or portability''\end{tabular}                                      \\
\textbf{\textit{SE\_E}}                & \begin{tabular}[c]{@{}l@{}}Word embedding\\(selected from top 50 words)~+ \\Original 2\end{tabular}          & \begin{tabular}[c]{@{}l@{}}``vulnerability, security, protection, cybersecurity, \\assurance, countermeasure, threat, privacy,\\~authentication, prevention, confidentiality, trusted,\\~intrusion, compromise, safety, insecure, defensive, \\breach, proactive, tampering, penetration, policy, phishing, \\vulnerable, authorization, dependability, or certification''\end{tabular} & \begin{tabular}[c]{@{}l@{}}``usability, performance, operational, \\look \& feel, legal, fault \& tolerance, \\maintainability, scalability, availability, or portability''\end{tabular}                                      \\
\bottomrule
\end{tabular}
}
\end{table}

\begin{table}
\centering

\caption{Label configurations for the top 4 largest NFR classes (US, SE, O, and PE) for the multi-class and multi-label classification sub-tasks in Task NFR.}
\label{tab:labelsNFRmulti}
\resizebox{1\textwidth}{!}{
\begin{tabular}{lll} 
\toprule
\vcell{\textbf{Label Abbr.}} & \vcell{\textbf{\textit{Label Configuration}}}                                            & \vcell{\textbf{\textit{List of Labels}}}                                                                                                                                                                                                                                                                                                                                                                                                                                                                                                                                                                                                                                                                                                                                                                                                                                                                                                                                                                                                                          \\[-\rowheight]
\printcelltop                 & \printcelltop                                                                             & \printcelltop                                                                                                                                                                                                                                                                                                                                                                                                                                                                                                                                                                                                                                                                                                                                                                                                                                                                                                                                                                                                                                                  \\ 
\hline
\vcell{\textit{MultiNFR\_A}}  & \vcell{Original label}                                                           & \vcell{{[}``usability'', ``security'', ``performance'', ``operational'']}                                                                                                                                                                                                                                                                                                                                                                                                                                                                                                                                                                                                                                                                                                                                                                                                                                                                                                                                                                                              \\[-\rowheight]
\printcelltop                 & \printcelltop                                                                             & \printcelltop                                                                                                                                                                                                                                                                                                                                                                                                                                                                                                                                                                                                                                                                                                                                                                                                                                                                                                                                                                                                                                                  \\
\vcell{\textit{MultiNFR\_B}}  & \vcell{Expert curated}                                                                  & \vcell{\begin{tabular}[b]{@{}l@{}}{[}``instructive, easy, helpful, useful, learnable, explainable, affordable, intuitive, or understandable'',\\``security, authorization, or protection'',\\``periodic execution or efficacy performance'',\\``working, running, connecting, interfacing, or operative environment'']\end{tabular}}                                                                                                                                                                                                                                                                                                                                                                                                                                                                                                                                                                                                                                                                                                                                   \\[-\rowheight]
\printcelltop                 & \printcelltop                                                                             & \printcelltop                                                                                                                                                                                                                                                                                                                                                                                                                                                                                                                                                                                                                                                                                                                                                                                                                                                                                                                                                                                                                                                  \\
\vcell{\textit{MultiNFR\_C}}  & \vcell{\begin{tabular}[b]{@{}l@{}}Word embedding\\~(selected from top 20 words)\end{tabular}} & \vcell{\begin{tabular}[b]{@{}l@{}}{[}``accessibility, aesthetic, contextual, experience, satisfaction, HCI, UX, \\questionnaire, ease, or ergonomics'' ,\\``vulnerability, securing. protecting, protection, cybersecurity, assurance, \\cyber, countermeasure, threat, privacy, authentication, prevention, or confidentiality'',\\``throughput, reliability, scalability, responsiveness, efficiency, workload, benchmark, \\latency, speed, improvement, or accuracy'',\\``environmental, organizational, coordination, systemic, or logistics'']\end{tabular}}                                                                                                                                                                                                                                                                                                                                                                                                                                                                                                     \\[-\rowheight]
\printcelltop                 & \printcelltop                                                                             & \printcelltop                                                                                                                                                                                                                                                                                                                                                                                                                                                                                                                                                                                                                                                                                                                                                                                                                                                                                                                                                                                                                                                  \\
\vcell{\textit{MultiNFR\_D}}  & \vcell{\begin{tabular}[b]{@{}l@{}}Word embedding\\(selected from top 50 words)\end{tabular}}  & \vcell{\begin{tabular}[b]{@{}l@{}}{[}``accessibility, aesthetic, contextual, experience, satisfaction, HCI, UX, questionnaire, ease, \\ergonomics, designer, evaluate, multimodal, practitioner, prototyping, preference, personalization, \\suitability, focus, clarity, responsiveness, judgement, feel, or helpful'',\\``vulnerability, security, protection, cybersecurity, assurance, countermeasure, threat, \\privacy, authentication, prevention, confidentiality, trusted, intrusion, compromise, safety,\\~insecure, defensive, breach, proactive, tampering, penetration, policy, phishing, vulnerable, \\authorization, dependability, or certification'',\\``throughput, reliability, scalability, responsiveness, efficiency, workload,\\~benchmark, latency, speed, improvement, accuracy, achieve, tuning, bottleneck,\\~better, high, optimize, effectiveness, low, enhances, reducing, increased, quality, faster, or degrades",\\``environmental, organizational, coordination, systemic, logistics, coordination, or automation'']\end{tabular}}  \\[-\rowheight]
\printcelltop                 & \printcelltop                                                                             & \printcelltop                                                                                                                                                                                                                                                                                                                                                                                                                                                                                                                                                                                                                                                                                                                                                                                                                                                                                                                                                                                                                                                  \\
\bottomrule
\end{tabular}
}
\end{table}

Two labelling configurations are used for the three sub-tasks, one for the binary case (Table~\ref{tab:labelsNFRbinary}), and the other for both the multi-class and multi-label classification cases (Table~\ref{tab:labelsNFRmulti}). 

\paragraph{Binary classification} For the binary case, we have 5 configurations for each NFR class considered, i.e., each binary ZSL classifier. In Table~\ref{tab:labelsNFRbinary} we report only the labels for the usability and security classes, while the other labels are reported in the 
online supplementary materials \footnote{\label{note1}\url{https://github.com/waadalhoshan/ZSL4REQ/tree/main/Appendix}}.
The strategies are analogous to those already discussed for the FR/NFR Task. The only main difference is the usage of the top-50 words from the word embeddings, besides the top-20. We performed some preliminary experiments and saw that the list of similar words for the NFR class names included relevant words also beyond the top-20, and therefore we considered it reasonable to extend the list of terms to be included in the labels.  This phenomenon was not observed for the previous task.

Concerning the agreement in the word selection (top-50) we have the following statistics: 52\% perfect, 44 \% partial, and 2\% disagreement. 
We obtained a Krippendorff's alpha rate of 0.45 and 0.53 at macro and micro level, respectively, 
and a Fleiss's kappa of 0.42 and 0.52 at macro and micro level, respectively 
(moderate agreement). 

\paragraph{Multi-class and multi-label classification} For these tasks, we have a list of labels for each configuration, cf. Table~\ref{tab:labelsNFRmulti}. The list is represented with squared brackets, the elements in the list are separated by commas, and each element is a label, expressed between quotes. In the table, each label in the list is associated to one of the top-4 largest classes, namely usability, security, performance and operational NFR. The label configurations considering \textit{all} classes are reported in 
GitHub\footnoteref{note1}.    
We did not use combinations of labelling strategies for these cases, given the extensive number of experiments, and the exploratory nature of the study.

\subsubsection{NFR Binary Classification}
Table \ref{tab:resultsdetailNFRbin} reports the classification results for the 4 largest NFR classes. The overall results indicate acceptable performance rates, with wF1 $> 0.71$ for all classes. 

The highest wF1 score of 0.84 is achieved for the security class, with AllMini + SE\_D, which uses the word embedding selected labels (top-20) for the security class, and the original NFR labels for the ``Other'' class. Following that is the usability class, with wF1 = 0.80, using Sbert + US\_E, which includes the word embedding selected labels (top-50) for the usability class, and the original labels for the ``Other'' class. This suggests that, for this task and for highly represented classes such as security and usability, generic LMs combined with word-embedding terms as labels appear to be the most effective configuration. It is also worth noting that this binary classification task leads to better results with respect to the FR vs NFR task (best wF1 = 84 vs wF1 =0.66), and the best results are obtained with more complex label configurations. This means that, when using ZSL, it is preferable to select relevant NFR classes and perform binary classification on them, rather than classifying FRs vs NFRs. Furthermore, while for the task FR/NFR simpler label selection strategies are preferable, more complex labels are appropriate for the NFR binary task.



\begin{table}
\centering
\scriptsize
\caption{Binary classification results of the top 4 NFR classes in \textit{Task NFR}.}
\label{tab:resultsdetailNFRbin}

 \resizebox{1\textwidth}{!}{
  
\begin{tabular}{l|l|c|c|c|ccc|ccc} 
\toprule


\multicolumn{1}{c|}{\multirow{2}{*}{\textbf{NFR class (249)}}} & \multicolumn{1}{c|}{\multirow{2}{*}{\textbf{ZSL classifier}}} 
& \multicolumn{3}{c|}{\textbf{Total}}
& \multicolumn{3}{c|}{\textbf{NFR class}}       & \multicolumn{3}{c}{\textbf{``Other'' class}}    \\ 
\cline{3-11}
\multicolumn{1}{c|}{}                                         & \multicolumn{1}{c|}{}      & \textbf{wP} & \textbf{wR} & \textbf{wF1}                                   & \textbf{P} & \textbf{R} & \textbf{F1} & \textbf{P} & \textbf{R}
& \textbf{F1}       \\ 
\hline

\textbf{US (67)}                                              & Sbert + US\_E                                                 & 0.81 & 0.82 & 0.80 &\textbf{0.73}  & 0.49          & 0.59           & 0.83           & \textbf{0.93} & \textbf{0.88}    \\
\textbf{SE (66)}                                              & AllMini + SE\_D                                                & \textbf{0.84} & \textbf{0.84} & \textbf{0.84} &0.67           & \textbf{0.73} & \textbf{0.70}  & \textbf{0.90}  & 0.87          & \textbf{0.89}       \\
\textbf{O (62)}                                               & Bert4RE + O\_C                                                & 0.72 & 0.73 & 0.72 & 0.46           & 0.37          & 0.41           & 0.80           & 0.86          & 0.83               \\
\textbf{PE (54)}                                              & Sbert + PE\_E                                                 & 0.78 & 0.78 & 0.78 & 0.50           & 0.46          & 0.48           & 0.85           & 0.87          & 0.86               \\
\textbf{PE (54)}                                              & AllMini + PE\_B                                               & 0.80 & 0.70 & 0.78 & 0.47           & 0.63          & 0.54           & 0.89           & 0.81          & 0.84               \\
\bottomrule
\end{tabular}
}
\end{table}

Table \ref{tab:resultsdetailNFRbin}, last six columns, considers P, R, and F1 for each class. We see that all the best classifiers tend to achieve higher performance on the ``Other'' class (best F1 for NFR class 0.70, vs 0.89 for the ``Other'' class). This suggests that the ZSL binary classifier encounters some difficulty in associating the requirements to the specific labels, despite the extensive set of terms used. A more accurate selection of terms, or the usage of terms directly coming from the requirements themselves\footnote{We did not consider this option, as it would have biased the classification. However, it is a viable choice in practical contexts.}, could overcome this issue. 



\begin{table}
\centering
\caption{Best performance results of binary classification of all 10 NFR classes in \textit{Task NFR.} The results are different with respect to the binary classification using solely four classes because more requirements are involved in the classification, thus leading to lower performance.}
  \label{tab:bestZSL4NFRbinary}
    
  \resizebox{1\textwidth}{!}{
  
\begin{tabular}{l|l|c|c|c|ccc|ccc} 
\toprule


\multicolumn{1}{c|}{\multirow{2}{*}{\textbf{NFR class (369)}}} & \multicolumn{1}{c|}{\multirow{2}{*}{\textbf{ZSL classifier}}} 
& \multicolumn{3}{c|}{\textbf{Total}}
& \multicolumn{3}{c|}{\textbf{NFR class}}       & \multicolumn{3}{c}{\textbf{``Other'' class}}    \\ 
\cline{3-11}
\multicolumn{1}{c|}{}                                         & \multicolumn{1}{c|}{}      & \textbf{wP} & \textbf{wR} & \textbf{wF1}                                   & \textbf{P} & \textbf{R} & \textbf{F1} & \textbf{P} & \textbf{R}
& \textbf{F1}       \\ 
\hline

US (67)                                                       & Sbert + US\_E                                             & 0.80                            & 0.79                           & 0.80                            & 0.44           & 0.49          & 0.46                               & 0.88           & 0.86          & 0.87                                   \\
SE (66)                                                       & AllMini + SE\_D                                           & 0.87                            & 0.85                           & 0.85                            & 0.61           & 0.33          & 0.43                               & 0.87           & 0.95          & 0.91                                   \\
O (62)                                                        & Bert4RE + O\_C                                            & 0.78                            & 0.77                           & 0.77                            & 0.32           & 0.37          & 0.35                               & 0.87           & 0.85          & 0.86                                   \\
PE (54)                                                       & Sbert + PE\_E                                             & 0.82                            & 0.78                           & 0.80                             & 0.33           & 0.46          & 0.38                               & 0.90           & 0.84          & 0.87                                   \\
LF (38)                                                       & Sbert + LF\_D                                             & 0.85                            & 0.76                           & 0.80                             & 0.18           & 0.21          & 0.20                               & 0.91           & 0.89          & 0.90                                   \\
A (21)                                                        & SObert + A\_D                                             & 0.90                            & 0.70                           & 0.78                            & 0.10           & 0.43          & 0.14                               & 0.95           & 0.72          & 0.82                                   \\
SC (21)                                                       & AllMini + SC\_E                                           & 0.92                            & 0.74                           & 0.81                            & 0.13           & 0.62          & 0.21                               & 0.97           & 0.75          & 0.85                                   \\
MN (17)                                                       & AllMini + MN\_E                                           & 0.93                            & 0.86                           & 0.89                            & 0.16           & 0.47          & 0.24                               & 0.97           & 0.88          & 0.92                                   \\
L (13)                                                        & AllMini + L\_E                                            & 0.95                            & 0.84                           & 0.88                            & 0.11           & 0.54          & 0.19                               & 0.98           & 0.85          & 0.91                                   \\
FT (10)                                                       & AllMini + FT\_E                                           & 0.96                            & 0.86                           & 0.91                            & 0.10           & 0.50          & 0.17                               & 0.98           & 0.88          & 0.93                                   \\
\bottomrule
\end{tabular}
}
\end{table}
Finally, Table ~\ref{tab:bestZSL4NFRbinary} lists the top performance rates considering all classes, and the entire set of requirements. Comparing these results with Table~\ref{tab:resultsdetailNFRbin}, we see that there is no substantial decrease in terms of performance for the largest classes, e.g., US still achieves wF1 = 0.80, while SE achieves wF1 = 0.85, which is even higher than wF1 = 0.84 in Table~\ref{tab:resultsdetailNFRbin}. However, the results are \textit{all} biased towards the `Other'' class, since, even in the best case, F1 for the NFR class is lower than 0.50. This problem was not so evident in Table~\ref{tab:resultsdetailNFRbin},
at least for the US and SE classes, where F1 is still acceptable also for the NFR class. We can therefore conclude that, in the case of  requirements belonging to many different classes, a binary ZSL classification leads to poor classification results, with the selected labelling strategies. 

\subsubsection{NFR Multi-class Classification}

\begin{table}
\centering
\caption{Multi-class classification results for top 4 NFR classes.} 
\label{tab:TopNFRmultiesults}
\resizebox{1\textwidth}{!}{
\begin{tabular}{l|ccc|ccc|ccc|ccc} 
\toprule
\multirow{2}{*}{\textbf{ZSL Classifier}} & \multicolumn{3}{c|}{\textbf{US}}                                                & \multicolumn{3}{c|}{\textbf{SE}}                                                & \multicolumn{3}{c|}{\textbf{O}}                                                 & \multicolumn{3}{c}{\textbf{PE}}                                                          \\ 
\cline{2-13}
                                         & \textbf{P}               & \textbf{R}               & \textbf{F1}               & \textbf{P}               & \textbf{R}               & \textbf{F1}               & \textbf{P}               & \textbf{R}               & \textbf{F1}               & \textbf{P}               & \textbf{R}               & \textbf{F1}                        \\ 
\hline
Sbert + MultiNFR\_A                      & 0.44                     & \textbf{0.52}            & 0.48                      & 0.57                     & 0.70                     & 0.63                      & 0.47                     & 0.45                     & 0.46                      & 0.43                     & 0.22                     & 0.29                               \\
Sbert + MultiNFR\_B                      & \textbf{0.73}            & 0.43                     & \textbf{\uline{0.54}}     & \textbf{\uline{0.69}}    & \textbf{0.64}            & \textbf{0.66}             & 0.53                     & \textbf{\uline{0.71}}    & \textbf{0.61}             & \textbf{0.48}            & 0.57                     & 0.52                               \\
Sbert + MultiNFR\_C                      & 0.62                     & 0.31                     & 0.42                      & 0.54                     & 0.62                     & 0.58                      & \textbf{0.59}            & 0.21                     & 0.31                      & 0.34                     & \textbf{\uline{0.74}}    & 0.47                               \\
Sbert + MultiNFR\_D                      & \multicolumn{1}{l}{0.69} & \multicolumn{1}{l}{0.40} & \multicolumn{1}{l|}{0.51} & \multicolumn{1}{l}{0.65} & \multicolumn{1}{l}{0.53} & \multicolumn{1}{l|}{0.58} & \multicolumn{1}{l}{0.38} & \multicolumn{1}{l}{0.23} & \multicolumn{1}{l|}{0.47} & \multicolumn{1}{l}{0.46} & \multicolumn{1}{l}{0.53} & \multicolumn{1}{l}{\textbf{0.58}}  \\ 
\hline
AllMini + MultiNFR\_A                    & 0.67                     & 0.36                     & 0.47                      & 0.66                     & 0.92                     & 0.77                      & 0.33                     & 0.32                     & 0.33                      & 0.45                     & 0.50                     & 0.47                               \\
AllMini + MultiNFR\_B                    & \textbf{\uline{0.77}}    & 0.36                     & \textbf{0.49}             & \textbf{0.63}            & \textbf{\uline{0.94}}    & \textbf{\uline{0.76}}     & \textbf{\uline{0.64}}    & \textbf{0.64}            & \textbf{\uline{0.64}}     & \textbf{\uline{0.59}}    & \textbf{0.70}            & \textbf{\uline{0.64}}              \\
AllMini + MultiNFR\_C                    & 0.54                     & \textbf{0.45}            & \textbf{0.49}             & 0.76                     & 0.67                     & 0.71                      & 0.36                     & 0.37                     & 0.37                      & 0.41                     & 0.54                     & 0.47                               \\
AllMini + MultiNFR\_D                    & 0.34                     & 0.43                     & 0.38                      & 0.88                     & 0.33                     & 0.48                      & 0.25                     & 0.42                     & 0.31                      & 0.47                     & 0.28                     & 0.35                               \\ 
\hline
Bert4RE + MultiNFR\_A                    & 0.32                     & 0.09                     & 0.14                      & \textbf{0.18}            & \textbf{0.11}            & \textbf{0.13}             & \textbf{0.33}            & 0.02                     & 0.03                      & \textbf{0.19}            & \textbf{0.67}            & \textbf{0.30}                      \\
Bert4RE + MultiNFR\_B                    & \textbf{0.33}            & \textbf{0.60}            & \textbf{0.42}             & \textit{0.00}            & \textit{0.00}            & \textit{0.00}             & 0.28                     & 0.53                     & \textbf{0.37}             & 0.13                     & 0.02                     & 0.03                               \\
Bert4RE + MultiNFR\_C                    & \textbf{0.33}            & 0.06                     & 0.10                      & 0.18                     & 0.08                     & 0.11                      & 0.24                     & \textbf{0.68}            & 0.36                      & 0.25                     & 0.17                     & 0.20                               \\
Bert4RE + MultiNFR\_D                    & \textit{0.00}            & \textit{0.00}            & \textit{0.00}             & \textit{0.00}            & \textit{0.00}            & \textit{0.00}             & 0.24                     & 0.34                     & 0.28                      & 0.25                     & 0.04                     & 0.06                               \\ 
\hline
SObert + MultiNFR\_A                     & \textit{0.00}            & \textit{0.00}            & \textit{0.00}             & 0.60                     & 0.05                     & 0.08                      & 0.22                     & 0.71                     & 0.33                      & \textbf{0.12}            & \textbf{0.09}            & \textbf{0.10}                      \\
SObert + MultiNFR\_B                     & 0.30                     & \textbf{\uline{0.81}}    & \textbf{0.44}             & \textit{0.00}            & \textit{0.00}            & \textit{0.00}             & 0.29                     & 0.16                     & 0.21                      & 0.08                     & 0.06                     & 0.07                               \\
SObert + MultiNFR\_C                     & \textbf{0.31}            & 0.36                     & 0.33                      & \textbf{0.24}            & \textbf{0.21}            & \textbf{0.22}             & \textbf{0.31}            & 0.56                     & \textbf{0.40}             & \textit{0.00}            & \textit{0.00}            & \textit{0.00}                      \\
SObert + MultiNFR\_D                     & 0.20                     & 0.04                     & 0.07                      & \textit{0.00}            & \textit{0.00}            & \textit{0.00}             & 0.24                     & \textbf{0.90}            & 0.38                      & \textit{0.00}            & \textit{0.00}            & \textit{0.00}                      \\
\bottomrule
\end{tabular}
}
\end{table}

Table \ref{tab:TopNFRmultiesults} reports the multi-class classification results for the 4 largest NFR classes. For this case, and for the multi-label case, we do not report the weighted measures for the sake of space. However, we remark that the results by class enable a more fine-grained analysis. Weighted F1 is presented in Table~\ref{tab:compareNFR} to facilitate comparison with other works. We see that, compared to the binary classification, results are substantially lower, although still acceptable for SE (F1 = 0.76), O (F1 = 0.64) and PE (F1 = 0.64). In the majority of the cases, the best results are obtained again with the domain-generic LMs, and using MultiNFR\_A or MultiNFR\_B as labels. These are the shortest labels, which do not use the word embedding strategies. This result is the opposite of what was observed for binary classification for NFR. We argue therefore that, in a multi-class classification setting, longer and more informative labels can lead to some possible overlapping between the represented meaning of each class. Instead, in a binary classification setting, more informative labels, i.e., using word embeddings, appear to be more effective. 


For the multi-class classification results for all the NFR classes (Table reported in supplementary material), the performance in terms of F1 remains acceptable only for SE class (F1= 0.69), while for the other classes poor results are obtained.

\subsubsection{NFR Multi-label Classification}

\begin{table}
\centering
\setlength{\extrarowheight}{0pt}
\addtolength{\extrarowheight}{\aboverulesep}
\addtolength{\extrarowheight}{\belowrulesep}
\setlength{\aboverulesep}{0pt}
\setlength{\belowrulesep}{0pt}
\caption{Multi-label classification results for 4 largest NFR classes---Top 2 results---249 requirements from the PROMISE dataset.}
\label{tab:TopNFRmultiLabelResults}
  \resizebox{1\textwidth}{!}{
\begin{tabular}{l|ccc|ccc|ccc|ccc} 
\toprule
\multicolumn{1}{c|}{\multirow{2}{*}{\textbf{ZSL Classifier}}} & \multicolumn{3}{c|}{\textbf{US}}                                      & \multicolumn{3}{c|}{\textbf{SE}}                                      & \multicolumn{3}{c|}{\textbf{O}}                                       & \multicolumn{3}{c}{\textbf{PE}}                                        \\ 
\cline{2-13}
\multicolumn{1}{c|}{}                                     & \textbf{P}        & \textbf{R}         & \textbf{F1}        & \textbf{P}        & \textbf{R}         & \textbf{F1}        & \textbf{P}        & \textbf{R}         & \textbf{F1}        & \textbf{P}        & \textbf{R}         & \textbf{F1}         \\ 
\hline
Sbert + MultiNFR\_A     & 0.60                  & 0.67                  & 0.60                  & 0.74                  & 0.84                  & 0.79                  & 0.72                  & 0.71                  & 0.72                  & 0.68                  & 0.46                  & 0.55                   \\
Sbert + MultiNFR\_B     & 0.93                  & 0.60                  & 0.73                  & \textbf{0.86}         & \textbf{0.77}         & \textbf{0.82}         & \textbf{0.71}         & \textbf{\uline{0.92}} & \textbf{0.80}         & \textbf{0.73}         & \textbf{0.91}         & \textbf{0.81}          \\
Sbert + MultiNFR\_C     & 0.84                  & 0.73                  & 0.78                  & 0.76                  & 0.79                  & 0.78                  & 0.81                  & 0.48                  & 0.60                  & 0.57                  & 0.91                  & 0.70                   \\
Sbert + MultiNFR\_D     & \textbf{0.92}         & \textbf{0.70}         & \textbf{0.80}         & 0.83                  & 0.76                  & 0.79                  & 0.62                  & 0.85                  & 0.72                  & 0.81                  & 0.80                  & 0.80                   \\
\hline
AllMini + MultiNFR\_A                                     & 0.82                  & 0.60                  & 0.69                  & 0.79                  & 1.00                  & 0.88                  & 0.62                  & 0.58                  & 0.60                  & 0.69                  & 0.74                  & 0.71                   \\
AllMini + MultiNFR\_B                                     & 0.93                  & 0.58                  & 0.72                  & \textbf{0.81}         & \textbf{\uline{0.98}} & \textbf{\uline{0.89}} & \textbf{0.80}         & \textbf{0.84}         & \textbf{0.82}         & \textbf{0.82}         & \textbf{\uline{0.94}} & \textbf{0.88}          \\
AllMini + MultiNFR\_C                                     & 0.63                  & 0.70                  & 0.66                  & 0.94                  & 0.73                  & 0.82                  & 0.63                  & 0.63                  & 0.63                  & 0.64                  & 0.72                  & 0.68                   \\
AllMini + MultiNFR\_D                                     & \textbf{0.63}         & \textbf{0.88}         & \textbf{0.74}         & 0.96                  & 0.74                  & 0.84                  & 0.61                  & 0.58                  & 0.60                  & 0.74                  & 0.63                  & 0.68                   \\
\hline
Bert4RE + MultiNFR\_A   & 0.84                  & 0.70                  & 0.76                  & \textbf{0.39}         & \textbf{0.20}         & \textbf{0.26}         & \textbf{0.96}         & \textbf{0.44}         & \textbf{0.60}         & 0.32                  & 0.78                  & 0.45                   \\
Bert4RE + MultiNFR\_B   & 0.41                  & 0.67                  & 0.51                  & 1.00                  & 0.08                  & 0.14                  & \textbf{0.45}         & \textbf{0.87}         & \textbf{0.60}         & \uline{0.93}          & 0.24                  & 0.38                   \\
Bert4RE + MultiNFR\_C   & \textbf{\uline{0.98}} & \textbf{0.76}         & \textbf{0.86}         & 0.28                  & 0.11                  & 0.15                  & 0.30                  & 0.73                  & 0.45                  & 0.53                  & 0.31                  & 0.40                   \\
Bert4RE + MultiNFR\_D   & 1.00                  & 0.12                  & 0.21                  & \textit{0.00}           & \textit{0.00}           & \textit{0.00}           & 0.31                  & 0.95                  & 0.47                  & \textbf{\uline{0.90}} & \textbf{0.87}         & \textbf{\uline{0.89}}  \\
\hline
SObert + MultiNFR\_A                                      & \textit{0.00}                     & \textit{0.00}                     & \textit{0.00}                     & \textbf{\uline{1.00}} & \textbf{0.44}         & \textbf{0.61}         & 0.36                  & 0.98                  & 0.53                  & \textbf{0.57}         & \textbf{0.54}         & \textbf{0.55}          \\
SObert + MultiNFR\_B                                      & 0.39                  & 0.85                  & 0.54                  & 1.00                  & 0.05                  & 0.09                  & \textbf{\uline{0.75}} & \textbf{\uline{0.92}} & \textbf{\uline{0.83}} & 0.29                  & 0.13                  & 0.18                   \\
SObert + MultiNFR\_C                                      & 0.71                  & 0.10                  & 0.83                  & 0.46                  & 0.38                  & 0.42                  & 0.43                  & 0.66                  & 0.52                  & 1.00                  & 0.09                  & 0.17                   \\
SObert + MultiNFR\_E                                      & \textbf{0.89}         & \textbf{\uline{1.00}} & \textbf{\uline{0.94}} 
& \textit{0.00}                     & \textit{0.00}                     & \textit{0.00}                     & 0.34                  & 0.92                  & 0.49                  & 1.00                  & 0.06                  & 0.11                   \\
\bottomrule
\end{tabular}
}
\end{table}
Table \ref{tab:TopNFRmultiLabelResults} reports the multi-label classification results for the 4 largest NFR classes, considering the top-2 labels returned by the classifier. In other terms, when the right label is returned by the classifier in the top-2 labels, we consider it a true positive. We see that in this case the performance  substantially increases with respect to the multi-class classification case in Table~\ref{tab:TopNFRmultiesults}, e.g., reaching $F1= 0.94$ for US requirements,  $0.89$ for SE, $0.83$ for O, and $0.89$ for PE. This suggests that the multi-label classification strategy may be the most effective when dealing with NFR classification. 

Looking at the results based on the LMs, we do not have a clear pattern, and each LM appears to be suitable for a certain requirement type. Concerning labels, simple configurations as MultiNFR\_A and MultiNFR\_B appears to be the most effective for all classes, except US, for which the embedding-based labels are more effective. This could be due to a better, and more clear-cut characterisation of US requirements with respect to other types. 

Performance is similar in case of the multi-label classification results for all the NFR classes (Table reported in supplementary material), considering the top-3 results---i.e., if the right label is returned among the top-3 labels, we consider it a true positive. The performance remain rather high, frequently with F1 above $0.90$ for the best configurations. 
Overall, we can say that ZSL in the multi-label classification context appears to be effective also in case of NFRs belonging to many classes.

\subsection{Task Security}

\subsubsection{Security Label Configuration}

\begin{table}
\centering
\caption{Label configurations for \textit{Task Security}.}
\label{tab:labelsSecurity}
 \resizebox{1\textwidth}{!}{
\begin{tabular}{llll} 
\toprule
\textbf{Label Abbr.} & \textbf{Label Configuration}                                                           & \textbf{Security}                                                                                                                                                                                                                                                                                                                                                                                   & \textbf{Non-Security}                                                                                                                                                                                                                                                                                                                                                                                          \\ 
\hline
Sec\_A               & Original label                                                                         & ``Security”                                                                                                                                                                                                                                                                                                                                                                                          & ``not about security”                                                                                                                                                                                                                                                                                                                                                                                           \\ 
\hline
Sec\_B               & Expert curated                                                                         & ``Security, authorization, or protection”                                                                                                                                                                                                                                                                                                                                                            & ``not about security, authorization, or protection”                                                                                                                                                                                                                                                                                                                                                             \\
\hline
Sec\_C               & \begin{tabular}[c]{@{}l@{}}Word embedding \\ (selected from top 20 words)\end{tabular} & \begin{tabular}[c]{@{}l@{}}``vulnerability, securing. protecting, \\ protection, cybersecurity, assurance, \\ cyber, countermeasure, \\ threat, privacy, authentication, \\ prevention, or confidentiality”\end{tabular}                                                                                                                                                                             & \begin{tabular}[c]{@{}l@{}}``not about vulnerability, securing. protecting, \\ protection, cybersecurity, assurance, cyber, \\ countermeasure, threat, privacy, authentication, \\ prevention, or confidentiality”\end{tabular}                                                                                                                                                                                 \\ 
\hline
Sec\_D               & \begin{tabular}[c]{@{}l@{}}Word embedding \\ (selected from top 50 words)\end{tabular} & \begin{tabular}[c]{@{}l@{}}``vulnerability, security, protection, \\ cybersecurity, assurance, countermeasure, \\ threat, privacy, authentication, \\ prevention, confidentiality, trusted, \\ intrusion, compromise, safety, insecure, \\ defensive, breach, proactive, tampering, \\ penetration, policy, phishing, \\ vulnerable, authorization, dependability, \\ or certification”\end{tabular} & \begin{tabular}[c]{@{}l@{}}``not about vulnerability, security, protection, \\ cybersecurity, assurance, countermeasure, \\ threat, privacy, authentication, \\ prevention, confidentiality, \\ trusted, intrusion, compromise, safety, \\ insecure, defensive, breach, \\ proactive, tampering, penetration, \\ policy, phishing, vulnerable, \\ authorization, dependability, or certification”\end{tabular}  \\
\bottomrule
\end{tabular}
}
\end{table}



The labelling of the security class (Table~\ref{tab:labelsSecurity}) is similar to the labels groups related to security as an NFR class in the binary classification task (cf. \ref{sec:NFRlabels}). The agreement results obtained for the word-embedding of the term ``security'' are the following:  50\% perfect, 48 \% partial, and 2\% disagreement. For IRR we obtained 0.46 as Krippendorff's alpha and a Fleiss' kappa score of 0.45, indicating a moderate agreement.



\subsubsection{Security Binary Classification}

Table~\ref{tab:resultsdetailSecurity}, column Total, reports the results for the Security task, considering all the requirements in the three datasets. The best performance is achieved by AllMini + Sec\_B, with a wF1 score of 0.66, with wP = 0.68 and wR = 0.65. The generic LM, AllMini, thus achieves best results. On the other hand, the other generic model, Sbert, achieves the worst results (wF1 = 0.31), thus suggesting that generic models are not necessarily better for this specific task. 
The best set of labels, Sec\_B, is the expert's curated one, which includes a limited set of three security-related words. This suggests that a limited number of well-selected terms is sufficient to identify security requirements in this dataset.

\begin{table}
\centering
\scriptsize
\caption{Binary classification results for \textit{Task Security}. All the requirements are considered together in one single SeqReq dataset.}
\label{tab:resultsdetailSecurity}

\begin{tabular}{l|ccc|ccc|ccc} 
\toprule
\multirow{2}{*}{\textbf{ZSL classifier}} & \multicolumn{3}{l|}{\textbf{Total}}   & \multicolumn{3}{l|}{ \textbf{Security (187)} } & \multicolumn{3}{l}{\textbf{Non-Security (323)}}\\ 
\cline{2-10}
                                     & \textbf{wP} & \textbf{wR} & \textbf{wF1} & \textbf{P} & \textbf{R} & \textbf{F1} & \textbf{P} & \textbf{R} & \textbf{F1} \\ 
\hline
Sbert + Sec\_C                         & 0.58 & 0.41  & 0.31& 0.37           & \textbf{0.92} & 0.53          & 0.70           & 0.11      & 0.19                \\
\textbf{AllMini + Sec\_B}            & \textbf{0.68} & \textbf{0.65} & \textbf{0.68} & \textbf{0.52}  & 0.67          & \textbf{0.58} & \textbf{0.77}  & 0.63          & \textbf{0.70}          \\
Bert4RE +  Sec\_A                     & 0.52 & 0.53 & 0.52 & 0.34           & 0.30          & 0.32          & 0.62           & \textbf{0.66} & 0.64                  \\
SObert +  Sec\_C                     & 0.56 & 0.54 & 0.55 & 0.39           & 0.51          & 0.45          & 0.66           & 0.55          & 0.60                  \\
\bottomrule
\end{tabular}

\end{table}

Table~\ref{tab:resultsdetailSecurity}, last six columns, shows the performance for the two classes. We see that better performance in terms of F1 is achieved for Non-Security requirements (F1 = 0.70 vs 0.58), using AllMini + Sec\_B, i.e., the best configuration. Looking in more detail, the best recall (R = 0.92) is obtained by  Sbert + Sec\_D. Therefore, if one seeks for a better ability to identify security requirements, i.e., high recall on this set, this configuration---though having the worst overall performance---should be preferred. This is an important observation, since for many requirement  tasks, including this one, high recall is more important than high precision, as remarked by Berry~\cite{berry2021empirical}---if one searches for security requirements, then one wants as less false negatives as possible.

\begin{table}
\centering
\scriptsize
\caption{Binary classification results for \textit{Task Security}. The results are obtained from each of the three projects in SecReq - CPN, GPS and ePurse.}
\label{tab:resultssubsetSecurity}
\begin{tabular}{lccc} 
\toprule
\multicolumn{1}{c}{\textbf{ZSL Classifier}} & \textbf{wP} & \textbf{wR} & \textbf{wF1}  \\ 
\hline
\multicolumn{4}{c}{\textbf{CPN}}                                                              \\ 
\hline
Sbert + Sec\_C                          & 0.85             & 0.30            & 0.25           \\
\textbf{AllMini + Sec\_B}               & \textbf{0.79}    & \textbf{0.77}   & \textbf{0.78}  \\
Bert4RE + Sec\_B                        & 0.65             & 0.80            & 0.72           \\
SObert + Sec\_A                         & 0.65             & 0.80            & 0.72           \\
SObert~

+ Sec\_B                       & 0.65             & 0.80            & 0.72           \\ 
\hline
\multicolumn{4}{c}{\textbf{GPS}}                                                              \\ 
\hline
Sbert+ Sec\_D                           & 0.67             & 0.40            & 0.30           \\
AllMini~~~~+ Sec\_B                     & 0.62             & 0.61            & 0.61           \\
Bert4RE + Sec\_C                        & 0.59             & 0.53            & 0.54           \\
\textbf{SObert~ ~+ Sec\_C}              & \textbf{0.63}    & \textbf{0.63}   & \textbf{0.63}  \\ 
\hline
\multicolumn{4}{c}{\textbf{ePurse}}                                                           \\ 
\hline
Sbert + Sec\_D                          & 0.59             & 0.64            & 0.60           \\
\textbf{\textbf{AllMini~}+ Sec\_A}      & \textbf{0.69}    & \textbf{0.70}   & \textbf{0.69}  \\
Bert4RE + Sec\_A                        & 0.67             & 0.44            & 0.40           \\
SObert~

+ Sec\_D                       & 0.66             & 0.69            & 0.62           \\
\bottomrule
\end{tabular}

\end{table}
Table~\ref{tab:resultssubsetSecurity} reports the results for the Security task, divided by each dataset included in SecReq. Best results are achieved for CPN (wF1 = 0.78), while worst results are for GPS (wF1 = 0.63). This could be due to the specific characteristics of the datasets. In some cases, security and non-security requirements in GPS are expressed with very similar sentences and are likely to be classified similarly though they belong to different classes (e.g., class Security: \textit{The Load File Data Block Hash is used in the computation of the Load File Data Block Signature} vs Non-Security: 
\textit{The Load File Data Block Hash is used in the computation of The Load Token}). 

\section{Research Findings}
\label{sec:find}

Based on the above detailed analysis of the experimental results, in this section we answer our three RQs one by one. Additionally, we also offer some general observations based on our experiments.

\subsection{Best Language Model (RQ1)}



\begin{itemize}
    \item For \textit{Task FR/NFR}, the best overall performance is obtained with Sbert, which has achieved a wF1 score of 0.66 (with wP = 0.71 and wR = 0.66). This indicates that the generic Sbert model, although designed to provide a semantic representation for generic sentences, substantially outperforms the other LMs (including two domain-specific LMs) for this task.
    
    \item For \textit{Task NFR}, the performance of LMs in each sub-task are as follows: 
    
    \begin{itemize}
    \item For binary classification of NFR, the generic AllMini model outperforms the other three LMs, particularly on the SE class. 
    \item For multi-class classification of NFR, in the majority of the cases, the best results are obtained by the generic LMs (Sbert and AllMini). 
    \item For multi-label classification of NFR, there is no clear winner as each LM appears to be suitable for a certain requirement class.
  \end{itemize}
  
    \item For \textit{Task Security}, the best overall performer is the generic AllMini model; on the other hand, the generic Sbert model achieves the worst results (wF1 = 0.31). This suggests that generic models are not necessarily better for this specific task and a careful selection of the best LM is key to the success of ZSL.
\end{itemize}

Based on the above findings, we can state that: 
\begin{mdframed}
\faLightbulbO{} In the majority of the cases, generic LMs perform better than domain-specific LMs on requirements classification tasks. When applying ZSL in practice one does not need to define domain- or project-specific LMs and can rely on larger ones that are freely available.
\end{mdframed}

Our findings thus contrast the claims that generic LMs do not perform particularly well on domain-specific tasks, as they cannot recognize highly domain-specific vocabulary \cite{beltagy2019scibert,chalkidis2020legal,lee2020biobert,sainani2020extracting,ajagbe2022RE}.  

Based on our experimental results we can conclude that \textit{generic LMs, being trained on generic data, are more generalizable and adaptable}---the actual sense of being generic; by contrast, \textit{domain-specific LMs, being trained on domain-specific data, are less generalizable and adaptable},---the actual sense of being specific. 
Future developments of LMs, we posit, should not differentiate between generic vs. specific, but rather, should focus on continual learning on new tasks and new data \cite{ruder2019transfer}. As LMs retain and accumulate knowledge across many tasks, they will become more adaptable to new tasks, domain-specific or otherwise. 

\subsection{Best Label Configuration (RQ2)}

\begin{itemize}
    \item For the \textit{Task FR/NFR}, the best label configuration is \textit{FR\_E} for Sbert. This configuration is composed of the Expert Curated and the Original labels, which identify the NFRs using the names of the NFR classes (Usability, Security, Availability, etc.). The result shows that expert knowledge of NFR characteristics plays an important role on label configuration for this task.
    
    \item For \textit{Task NFR}, we show the performance of label configurations for each sub-task as follows: 
    
    \begin{itemize}
    \item For binary classification of NFR, the best label configuration is \textit{SE\_D} for AllMini, which uses the word embedding with top-20 words for the SE class, and the original NFR labels for the ``Other'' class. 
    \item For multi-class classification of NFR, in the majority of the cases, the best label configurations for individual NFR classes are \textit{MultiNFR\_A} (Original label) and \textit{MultiNFR\_B} (Expert curated label) for Sbert and AllMini. 
    \item For multi-label classification of NFR, simple label configurations based on either original label (\textit{MultiNFR\_A})  and expert curated label (\textit{MultiNFR\_B}) appear to be most effective for all classes, except US, for which the embedding-based labels (\textit{MultiNFR\_D} and \textit{MultiNFR\_E}) are more effective. 
  \end{itemize}
    
    \item For \textit{Task Security}, the best label configuration is \textit{Sec\_B}, curated by expert. Although this label only contains three security-related words, it has shown to be effective in identifying security requirements. 
\end{itemize}

Based on the above findings, we can conclude that: 

\begin{mdframed}

\faLightbulbO{} Label selection has a relevant impact on the performance of ZSL classification. In general, simple label configurations with the original class names or with a combination of original and expert-curated labels appear to be more effective than more complex word-embedding generated labels. 
\end{mdframed}
\begin{mdframed}
\faLightbulbO{}
The above conclusion implies that, when applying ZSL in practice, for a given requirements classification scheme, domain experts can manually select their own labels for the classes, without using word embeddings. Furthermore, they can also consider project-specific terms that can better distinguish between classes. Preliminary trials to select the best configuration for the problem at hand are also recommended.
\end{mdframed}
\begin{mdframed}
\faLightbulbO{}

An exception is the binary classification of NFRs, where word-embeddings enable better performance. In these cases, more complex label configurations based on word embeddings should be preferred.

\end{mdframed}

Selecting the most effective label configuration is a difficult task and requires testing many different labels by trial and error. Our study shows how we have handcrafted each label using one of the aforementioned three strategies. 
However, more work is needed in search for a more systematic approach to label configuration. We argue that expert knowledge of RE, both domain-specific and possibly project-specific, plays an important part in choosing the correct terms for the labels. 

\subsection{Effectiveness of ZSL for RE (RQ3)}

Here we address the effectiveness of ZSL by first comparing our best ZSL results to the state-of-the-art results achieved by Kurtanovi\`c and Maleej (K\&M) \cite{kurtanovic2017automatically}, Hey \textit{et al.} (NoRBERT) \cite{hey2020norbert} and Knauss \textit{et al.} (Knauss) \cite{knauss2011supporting}, with respect to the same classification tasks (i.e., binary and multi-class classification). Second, we discuss our best ZSL results obtained from multi-label classification with state-of-the-art results and provide our insight into ZSL classification. 

\begin{table}
\centering
\caption{Binary classification results obtained from \textit{Task FR/NFR} compared to the results obtained by K\&M \cite{kurtanovic2017automatically} and NoRBERT \cite{hey2020norbert}. }
\label{tab:compareFRvsNFR}
\begin{tabular}{l|rrr|rrr|r} 
\toprule
\multicolumn{1}{c|}{\multirow{2}{*}{\textbf{ Approach (model, train/test)}}}                                                         & \multicolumn{3}{c|}{\textbf{FR (255)}}                                                             & \multicolumn{3}{c|}{\textbf{NFR (370)}}                                                            & \multicolumn{1}{c}{\multirow{2}{*}{\textbf{wF1}}}  \\ 
\cline{2-7}
\multicolumn{1}{c|}{}                                                                                                                & \multicolumn{1}{c}{\textbf{P}} & \multicolumn{1}{c}{\textbf{R}} & \multicolumn{1}{c|}{\textbf{F1}} & \multicolumn{1}{c}{\textbf{P}} & \multicolumn{1}{c}{\textbf{R}} & \multicolumn{1}{c|}{\textbf{F1}} & \multicolumn{1}{c}{}                               \\ 
\hline
K\&M (word features, 10-fold)                                                                                                          & \textbf{0.92}                  & 0.93                           & \textbf{0.93}                    & 0.93                           & 0.92                           & 0.92                             & 0.92                                               \\
K\&M (best 100 features, 10-fold)                                                                                                      & 0.86                           & 0.51                           & 0.63                             & 0.65                           & 0.92                           & 0.76                             & 0.71                                               \\
K\&M (best 500 features, 10-fold)                                                                                                      & \textbf{0.92}                  & 0.79                           & 0.85                             & 0.82                           & 0.93                           & 0.87                             & 0.86                                               \\ 
\hline
NoRBERT (base+ep.16 \tablefootnote{ep. refers to the number of passes of the training dataset during LM learning process.}, 10-fold) & 0.89                           & 0.88                           & 0.89                             & 0.92                           & 0.93                           & 0.92                             & 0.91                                               \\
NoRBERT (large+ep.10+OS, 10-fold)                                                                                                    & \textbf{0.92}                  & 0.88                           & 0.90                             & 0.92                           & \textbf{0.95}                  & 0.93                             & 0.92                                               \\ 
\hline
ZSL(Sbert + FR\_E, all)                                                                                                              & \textbf{\uline{0.55}}          & \textbf{\uline{0.82}}          & \textbf{\uline{0.66}}            & \textbf{\uline{0.82}}          & 0.54                           & 0.65                             & \textbf{\uline{0.65}}                              \\
ZSL(AllMini + FR\_D, all)                                                                                                            & 0.50                           & 0.71                           & 0.58                             & 0.72                           & 0.50                           & 0.59                             & 0.59                                               \\
ZSL(Bert4RE + FR\_D, all)                                                                                                            & 0.47                           & 0.65                           & 0.54                             & 0.67                           & 0.50                           & 0.57                             & 0.56                                               \\
ZSL(SObert + FR\_C, all)                                                                                                             & 0.50                           & 0.41                           & 0.45                             & 0.64                           & \textbf{\uline{0.72}}          & \textbf{\uline{0.68}}            & 0.59                                               \\
\bottomrule
\end{tabular}
\end{table}


\begin{table}
\centering
\caption{Binary and multi-class classification results obtained from \textit{Task NFR} compared to the results obtained by K\&M and NoRBERT. Only top 4 NFR classes are considered.}

\label{tab:compareNFR}
  \resizebox{1\textwidth}{!}{
\begin{tabular}{c|l|rrr|rrr|rrr|rrr|c} 
\toprule
\multicolumn{2}{c|}{\multirow{2}{*}{\begin{tabular}[c]{@{}c@{}}\textbf{Approach (parameters)}\\\end{tabular}}}                                                                                                                                                                                                  & \multicolumn{3}{c|}{\textbf{US (67)}}                                                              & \multicolumn{3}{c|}{\textbf{SE (66)}}                                                              & \multicolumn{3}{c|}{\textbf{O (62)}}                                                               & \multicolumn{3}{c|}{\textbf{PE (54)}}                                                              & \multirow{2}{*}{\textbf{wF1}}  \\ 
\cline{3-14}
\multicolumn{2}{c|}{}                                                                                                                                                                                                                                                                                           & \multicolumn{1}{c}{\textbf{P}} & \multicolumn{1}{c}{\textbf{R}} & \multicolumn{1}{c|}{\textbf{F1}} & \multicolumn{1}{c}{\textbf{P}} & \multicolumn{1}{c}{\textbf{R}} & \multicolumn{1}{c|}{\textbf{F1}} & \multicolumn{1}{c}{\textbf{P}} & \multicolumn{1}{c}{\textbf{R}} & \multicolumn{1}{c|}{\textbf{F1}} & \multicolumn{1}{c}{\textbf{P}} & \multicolumn{1}{c}{\textbf{R}} & \multicolumn{1}{c|}{\textbf{F1}} &                               \\ 
\hline
\multicolumn{15}{c}{\textbf{NFR Binary Classification~}}                                                                                                                                                                                                                                                                                                                                                                                                                                                                                                                                                                                                                                                                                                            \\ 
\hline
\multirow{6}{*}{\begin{tabular}[c]{@{}c@{}}10-fold\\cross val.\\on Top-4\\NFR (249)\end{tabular}}                 & K\&M(w/o features selection)                                                                                                                                                                  & 0.81                           & \textbf{0.85}                  & 0.82                             & 0.91                           & \textbf{0.90}                  & 0.88                             & 0.72                           & \textbf{0.75}                  & 0.73                             & \textbf{0.93}                  & \textbf{0.90}                  & \textbf{0.90}                    & \textbf{0.83}                 \\
                                                                                                                  & K\&M (best 50 features)                                                                                                                                                                        & 0.70                           & 0.57                           & 0.61                             & 0.81                           & 0.77                           & 0.74                             & 0.78                           & 0.50                           & 0.57                             & 0.87                           & 0.57                           & 0.67                             & 0.65                          \\
                                                                                                                  & K\&M (best 500 features)                                                                                                                                                                       & 0.80                           & 0.71                           & 0.74                             & 0.74                           & 0.81                           & 0.74                             & 0.72                           & 0.73                           & 0.71                             & 0.87                           & 0.81                           & 0.82                             & 0.75                          \\
                                                                                                                  & NoRBERT(base + ep.10)                                                                                                                                                                       & 0.81                           & 0.69                           & 0.74                             & \textbf{0.93}                  & 0.82                           & 0.87                             & 0.80                           & 0.53                           & 0.64                             & 0.88                           & 0.80                           & 0.83                             & 0.77                          \\
                                                                                                                  & NoRBERT(base + ep.10 - OS\tablefootnote{OS refers to Oversampling technique for randomly selecting data from the minority class by adding them to the training dataset Sampling })          & 0.78                           & 0.70                           & 0.74                             & 0.90                           & 0.86                           & 0.88                             & 0.88                           & 0.71                           & \textbf{0.79}                    & 0.88                           & 0.80                           & 0.83                             & 0.81                          \\
                                                                                                                  & NoRBERT(large+OS+ES\tablefootnote{ES refers Early Stopping, a feature that enables the model training to be automatically stopped when a selected metric (e.g., F1) has stopped improving}) & \textbf{0.89}                  & 0.70                           & \textbf{0.78}                    & 0.89                           & 0.89                           & \textbf{0.89}                    & \textbf{0.90}                  & 0.71                           & \textbf{0.79}                    & 0.88                           & 0.81                           & 0.85                             & \textbf{0.83}                 \\ 
\hline
\multicolumn{1}{l|}{\multirow{8}{*}{\begin{tabular}[c]{@{}l@{}}Test Top-4\\NFR (249)\\w/o training\end{tabular}}} & ZSL(Sbert + (NFR)\_D)                                                                                                                                                                       & 0.77                           & 0.78                           & 0.78                             & 0.72                           & 0.71                           & 0.72                             & 0.68                           & 0.72                           & 0.70                             & 0.80                           & 0.67                           & 0.70                             & 0.72                          \\
\multicolumn{1}{l|}{}                                                                                             & \textbf{ZSL(Sbert + (NFR)\_E)}                                                                                                                                                              & \textbf{\uline{0.81}}          & \textbf{\uline{0.82}}          & \textbf{\uline{0.80}}            & 0.76                           & 0.78                           & 0.75                             & 0.68                           & 0.63                           & 0.65                             & 0.78                           & \textbf{\uline{0.78}}          & \textbf{\uline{0.78}}            & \textbf{\uline{0.74}}         \\
\multicolumn{1}{l|}{}                                                                                             & ZSL(AllMini + (NFR)\_B)                                                                                                                                                                     & 0.54                           & 0.35                           & 0.35                             & 0.73                           & 0.73                           & 0.73                             & 0.71                           & 0.65                           & 0.67                             & 0.80                           & 0.70                           & \textbf{\uline{0.78}}            & 0.62                          \\
\multicolumn{1}{l|}{}                                                                                             & \textbf{ZSL(AllMini + (NFR)\_D)}                                                                                                                                                            & 0.75                           & 0.75                           & 0.75                             & \textbf{\uline{0.84}}          & \textbf{\uline{0.84}}          & \textbf{\uline{0.84}}            & 0.65                           & 0.55                           & 0.58                             & \textbf{\uline{0.81}}          & 0.69                           & 0.71                             & 0.72                          \\
\multicolumn{1}{l|}{}                                                                                             & \textbf{ZSL(BERT4RE + (NFR)\_C)}                                                                                                                                                            & 0.52                           & 0.43                           & 0.46                             & 0.62                           & 0.65                           & 0.63                             & \textbf{\uline{0.72}}          & \textbf{\uline{0.73}}          & \textbf{\uline{0.72}}            & 0.68                           & 0.41                           & 0.43                             & 0.56                          \\
\multicolumn{1}{l|}{}                                                                                             & ZSL(BERT4RE + (NFR)\_D)                                                                                                                                                                     & 0.54                           & 0.42                           & 0.45                             & 0.56                           & 0.70                           & 0.61                             & 0.63                           & 0.41                           & 0.42                             & 0.60                           & 0.46                           & 0.51                             & 0.50                          \\
\multicolumn{1}{l|}{}                                                                                             & ZSL(SObert + (NFR)\_B)                                                                                                                                                                      & 0.55                           & 0.51                           & 0.53                             & 0.81                           & 0.74                           & 0.63                             & 0.65                           & \textbf{\uline{0.73}}          & 0.67                             & 0.61                           & \textbf{\uline{0.78}}          & 0.68                             & 0.62                          \\
\multicolumn{1}{l|}{}                                                                                             & ZSL(SObert + (NFR)\_C)                                                                                                                                                                      & 0.51                           & 0.32                           & 0.30                             & 0.66                           & 0.53                           & 0.56                             & 0.71                           & 0.74                           & 0.71                             & 0.61                           & 0.55                           & 0.58                             & 0.53                          \\ 
\hline
\multicolumn{15}{c}{\textbf{NFR~Multi-class Classification~}}                                                                                                                                                                                                                                                                                                                                                                                                                                                                                                                                                                                                                                                                                                       \\ 
\hline
\multirow{5}{*}{\begin{tabular}[c]{@{}c@{}}10-fold\\cross val.\\on Top-4\\NFR (249)\end{tabular}}                 & K\&M(word features)                                                                                                                                                                           & 0.65                           & 0.82                           & 0.70                             & 0.81                           & 0.77                           & 0.75                             & 0.81                           & \textbf{0.86}                  & \textbf{0.82}                    & 0.86                           & \textbf{0.81}                  & 0.80                             & 0.76                          \\
                                                                                                                  & K\&M(best 50 features)                                                                                                                                                                        & 0.49                           & 0.68                           & 0.55                             & 0.60                           & 0.50                           & 0.39                             & 0.42                           & 0.47                           & 0.33                             & 0.85                           & 0.53                           & 0.63                             & 0.47                          \\
                                                                                                                  & K\&M(best 500 features)                                                                                                                                                                       & 0.70                           & 0.66                           & 0.64                             & 0.64                           & 0.53                           & 0.56                             & 0.47                           & 0.62                           & 0.51                             & 0.81                           & 0.74                           & 0.76                             & 0.61                          \\
                                                                                                                  & NoRBERT(base+ep.32)                                                                                                                                                                         & 0.78                           & \textbf{0.84}                  & 0.81                             & 0.89                           & 0.85                           & 0.87                             & 0.79                           & 0.73                           & 0.76                             & 0.88                           & 0.78                           & 0.82                             & 0.82                          \\
                                                                                                                  & NoRBERT(large+ep.32)                                                                                                                                                                        & \textbf{0.86}                  & 0.82                           & \textbf{0.84}                    & \textbf{0.91}                  & \textbf{0.91}                  & \textbf{0.91}                    & \textbf{0.83}                  & 0.71                           & 0.77                             & \textbf{0.90}                  & \textbf{0.81}                  & \textbf{0.85}                    & 0.84                          \\ 
\hline
\multirow{11}{*}{\begin{tabular}[c]{@{}c@{}}Test Top-4\\NFR (249)\\w/o training\end{tabular}}                     & \textbf{ZSL(Sbert + MultiNFR\_B)}                                                                                                                                                           & \textbf{0.73}                  & \textbf{0.43}                  & \uline{0.54}                     & 0.69                           & \textbf{0.64}                  & \textbf{0.66}                    & \textbf{0.53}                  & 0.71                           & \textbf{0.61}                    & \textbf{0.48}                  & \textbf{0.57}                  & \textbf{0.52}                    & \textbf{0.58}                 \\
                                                                                                                  & ZSL(Sbert + MultiNFR\_D)                                                                                                                                                                    & 0.69                           & 0.40                           & 0.51                             & 0.65                           & 0.53                           & 0.58                             & 0.38                           & 0.23                           & 0.47                             & 0.46                           & 0.53                           & 0.58                             & 0.53                          \\
                                                                                                                  & ZSL(AllMini + MultiNFR\_A)                                                                                                                                                                  & 0.67                           & 0.36                           & 0.47                             & 0.66                           & 0.92                           & 0.77                             & 0.33                           & 0.32                           & 0.33                             & 0.45                           & 0.50                           & 0.47                             & 0.51                          \\
                                                                                                                  & \textbf{ZSL(AllMini + MultiNFR\_B)}                                                                                                                                                         & \textbf{\uline{0.77}}          & 0.36                           & 0.49                             & 0.63                           & \textbf{\uline{0.94}}          & \textbf{\uline{0.76}}            & \textbf{\uline{0.64}}          & 0.64                           & \textbf{\uline{0.64}}            & \textbf{\uline{0.59}}          & \textbf{\uline{0.70}}          & \textbf{\uline{0.64}}            & \uline{\textbf{0.63}}         \\
                                                                                                                  & ZSL(AllMini + MultiNFR\_C)                                                                                                                                                                  & 0.54                           & 0.45                           & 0.49                             & 0.76                           & 0.67                           & 0.71                             & 0.36                           & 0.37                           & 0.37                             & 0.41                           & 0.54                           & 0.47                             & 0.51                          \\
                                                                                                                  & ZSL(BERT4RE + MultiNFR\_A)                                                                                                                                                                  & 0.32                           & 0.09                           & 0.14                             & 0.18                           & 0.11                           & 0.13                             & 0.33                           & 0.02                           & 0.03                             & 0.19                           & 0.67                           & 0.30                             & 0.14                          \\
                                                                                                                  & ZSL(BERT4RE + MultiNFR\_B)                                                                                                                                                                  & 0.33                           & 0.60                           & 0.42                             & \textit{0.00}                  & \textit{0.00}                  & \textit{0.00}                    & 0.28                           & 0.53                           & 0.37                             & 0.13                           & 0.02                           & 0.03                             & 0.21                          \\
                                                                                                                  & ZSL(BERT4RE + MultiNFR\_D)                                                                                                                                                                  & \textit{0.00}                  & \textit{0.00}                  & \textit{0.00}                    & \textit{0.00}                  & \textit{0.00}                  & \textit{0.00}                    & 0.24                           & 0.34                           & 0.38                             & 0.25                           & 0.04                           & 0.06                             & 0.11                          \\
                                                                                                                  & ZSL(SObert + MultiNFR\_A)                                                                                                                                                                   & \textit{0.00}                  & \textit{0.00}                  & \textit{0.00}                    & 0.60                           & 0.05                           & 0.08                             & 0.22                           & \textbf{\uline{0.71}}          & 0.33                             & 0.12                           & 0.09                           & 0.10                             & 0.13                          \\
                                                                                                                  & ZSL(SObert + MultiNFR\_B)                                                                                                                                                                   & 0.30                           & \textbf{\uline{0.81}}          & 0.44                             & \textit{0.00}                  & \textit{0.00}                  & \textit{0.00}                    & 0.29                           & 0.16                           & 0.21                             & 0.08                           & 0.06                           & 0.07                             & 0.19                          \\
                                                                                                                  & ZSL(SObert + MultiNFR\_C)                                                                                                                                                                   & 0.31                           & 0.36                           & 0.33                             & 0.24                           & 0.21                           & 0.22                             & 0.31                           & 0.56                           & 0.40                             & \textit{0.00}                  & \textit{0.00}                  & \textit{0.00}                    & 0.25                          \\
\bottomrule
\end{tabular}
}
\end{table}

\begin{table}
\centering

\caption{Comparison between the classification results obtained by Knauss et al. (2011) and \textit{Task Security} divided by the subset of SeqReq dataset: CPN, GPS, and ePurse. Bold values indicate the best performance results. Underlined values refer to the best performance rates with the ZSL classifier.}
\label{tab:compareSec}

\resizebox{1\textwidth}{!}{
\begin{tabular}{l|l|rrr} 
\toprule
\multicolumn{2}{c|}{\textbf{Approach (parameters)}}                                                                                & \multicolumn{1}{l}{\textbf{wP}} & \multicolumn{1}{l}{\textbf{wR}} & \multicolumn{1}{l}{\textbf{wF1}}  \\ 
\hline
\multicolumn{5}{c}{\textbf{SeqReq (510)}}                                                                                                                                                                                                  \\ 
\hline
\begin{tabular}[c]{@{}l@{}}10-fold cross val.\\on SeqReq (510)\end{tabular}               & Knauss et al. (Bayesian classifier)      & \textbf{0.79}                   & \textbf{0.91}                   & \textbf{0.84}                     \\
\multirow{4}{*}{\begin{tabular}[c]{@{}l@{}}Test SeqReq (510)~\\w/o training\end{tabular}} & ZSL(Sbert + Labels\_Sec\_C)            & 0.58                            & 0.41                            & 0.31                              \\
                                                                                          & \textbf{ZSL(AllMini + Labels\_Sec\_B)} & \textbf{\uline{0.68}}           & \textbf{\uline{0.65}}           & \textbf{\uline{0.66}}             \\
                                                                                          & ZSL(BERT4RE + Labels\_Sec\_A)          & 0.52                            & 0.53                            & 0.52                              \\
                                                                                          & ZSL(SObert + Labels\_Sec\_C)           & 0.56                            & 0.54                            & 0.55                              \\ 
\hline
\multicolumn{5}{c}{\textbf{CPN (210)}}                                                                                                                                                                                                     \\ 
\hline
Train on ePurse (124)                                                                     & Knauss et al. (Bayesian classifier)      & 0.23                            & 0.54                            & 0.33                              \\
Train on GPS (176)                                                                        & Knauss et al. (Bayesian classifier)      & \textbf{0.29}                   & 0.65                            & \textbf{0.40}                     \\
Train on~ePurse + GPS (300)                                                               & Knauss et al. (Bayesian classifier)      & 0.26                            & \textbf{0.85}                   & \textbf{0.40}                     \\
Test CPN (210)~w/o training                                                               & \textbf{ZSL(AllMini + Labels\_Sec\_B)} & \uline{\textbf{0.79}}           & \uline{\textbf{0.77}}           & \uline{\textbf{0.78}}             \\ 
\hline
\multicolumn{5}{c}{\textbf{GPS (176)}}                                                                                                                                                                                                     \\ 
\hline
Train on ePurse (124)                                                                     & Knauss et al. (Bayesian classifier)      & 0.43                            & \textbf{0.85}                   & \textbf{0.57}                     \\
Train on CPN (210)                                                                        & Knauss et al. (Bayesian classifier)      & 0.29                            & 0.19                            & 0.23                              \\
Train on ePurse + CPN (334)                                                               & Knauss et al. (Bayesian classifier)      & \textbf{0.51}                   & 0.56                            & 0.53                              \\
\begin{tabular}[c]{@{}l@{}}Test GPS (176) \\w/o training\end{tabular}                     & \textbf{ZSL(SObert + Labels\_Sec\_C)}  & \uline{\textbf{0.63}}           & \uline{\textbf{0.63}}           & \uline{\textbf{0.63}}             \\ 
\hline
\multicolumn{5}{c}{\textbf{ePurse (124)}}                                                                                                                                                                                                  \\ 
\hline
Train on CPN (210)                                                                        & Knauss et al. (Bayesian classifier)      & \textbf{0.99}                   & 0.33                            & 0.47                              \\
Train on GPS (176)                                                                        & Knauss et al. (Bayesian classifier)      & 0.72                            & \textbf{0.48}                   & \textbf{0.58}                     \\
Train on ePurse + CPN (386)                                                               & Knauss et al. (Bayesian classifier)      & 0.84                            & 0.31                            & 0.46                              \\
Test ePurse (124) w/o training                                                            & \textbf{ZSL(AllMini + Labels\_Sec\_A)} & \textbf{\uline{0.64}}           & \textbf{\uline{0.70}}           & \textbf{\uline{0.69}}             \\
\bottomrule
\end{tabular}
}
\end{table}

\begin{itemize}
    \item \textit{Binary Classification of FR vs. NFR:} Table \ref{tab:compareFRvsNFR} shows that both K\&M and NoRBERT outperform all our ZSL classifiers. In particular, on FR, K\&M produces the best results with a SVM model that applies all the word features in the PROMISE dataset (i.e., without feature selection), achieving F1 = 0.93. On NFR, NoRBERT produces the best results with the fine-tuned BERT\textsubscript{large} model, with F1 = 0.93. By contrast, the best ZSL classifier (with Sbert LM) has only managed to achieve F1 = 0.66 on FR and F1 = 0.65 on NFR. On average, the performance of the best ZSL classifier is 0.27 lower than that of K\&M and NoRBERT. Clearly, these results show that the ZSL approach is (much) less effective than K\&M and NoRBERT with respect to this particular task. 
    
    \item \textit{Binary Classification of NFRs}: Table \ref{tab:compareNFR} shows that overall, both K\&M and NoRBERT outperform our best ZSL classifier, with wF1 = 0.83 achieved by their best model; on the other hand, the performance of our best classifier (ZSL with Sbert) is 0.10 points worse, with wF1 = 0.73. By examining the results obtained for each class, on US, our ZSL classifier (with Sbert) performs slightly worse than K\&M, but outperforms NoRBERT. On SE, although both K\&M and NoRBERT outperform our best ZSL classifier (with AllMini), the difference is not large. A similar observation can be made to classes Operational (O) and Performance (PE).
    
    \item \textit{Multi-class Classification of NFRs}: For this task, we notice in Table \ref{tab:compareNFR}
    large gaps exist between the results of K\&M and NoRBERT and our results on every class. As the purpose of this task is basically the same as the binary classification of NFR task, the inconsistent results achieved by ZSL in these two tasks indicate that when a requirement belongs to many classes, ZSL does not appear to be sufficiently effective. 
    \item \textit{Binary Classification of Security vs Non-Security Requirements}: Table \ref{tab:compareSec} reveals interesting results. When treating all the security requirements as a whole (i.e., without separating them into different projects), Knauss outperforms the best ZSL classifier by 0.18 points on wF1. However, when the requirements are divided into three projects (i.e., CPN, GPS and ePurse), ZSL outperforms Knauss on all individual projects. In particular, ZSL (AllMini + Sec\_B) achieved a high wF1 = 0.78, compared to Knauss's wF1 = 0.40 on CPN. This again seems to suggest that ZSL performs well with binary classification of security requirements when opposite labels are clearly defined.
\end{itemize}

Based on the above findings, we can conclude that:

\begin{mdframed}
\faLightbulbO{} Unsupervised learning with ZSL achieves acceptable performance for binary and multi-class classification tasks. However, it does not outperform supervised classification models, as RE tasks are narrowly defined, and often require well-trained, specifically fine-tuned models on specifically labelled dataset. Nevertheless, without training or fine-tuning, ZSL is more flexible, open to less data-rich tasks, and easily adaptable to the evolution of classification schemes. 
\end{mdframed} 

\begin{mdframed}
\faLightbulbO{} When using ZSL in practice, a company can choose its requirements classification scheme. This will also entail selection of new labels. Given its lower performance in comparison with supervised methods, ZSL is recommended for contexts with large sets of non-mission critical requirements, where misclassification can be tolerated.
\end{mdframed}

In relation to multi-label classification, the following results are obtained: From Table~\ref{tab:TopNFRmultiLabelResults}, concerning the 4 largest NFR classes, best performance for each class are F1 $\sim 0.83-0.94$, which are comparable with the average results of NoRBERT (large + ep.32) for multi-class classification (average F1 = 0.84, cf. 
Table \ref{tab:compareNFR}), and are higher than those of K\&M.

\begin{mdframed}
\faLightbulbO{} To achieve state-of-the-art performance of ZSL for multi-class classification, a multi-label strategy is recommended. In practice, this implies that a semi-automated classification approach should be followed, in which a human operator is asked to select the most suitable class among the top ones returned by the ZSL classifier. 
\end{mdframed}

\section{Threats to Validity}
\label{sec:threats}

\paragraph{Construct Validity} The first threat in our study is the adopted concept of FR and NFRs. This is an artificial distinction~\cite{eckhardt2016non}, as NFRs are often referred to as \textit{qualities}~\cite{ieeestandard}, and their classification is often non-binary, i.e., a multi-label classification. However, FR/NFR is a traditional distinction, still common in industrial practice and research. Furthermore, using ZSL for a multi-label binary case would introduce the need for threshold values in the classification (i.e., when both classes have a correlation score above a certain threshold, then classify the requirement as \textit{both} FR and NFR). 
For this reason, we excluded the binary, multi-label variant of the PROMISE dataset annotated by Dalpiaz \textit{et al.}~\cite{dalpiaz2019requirements} from our evaluation. Therefore, the presented approach does not apply to cases in which a requirement can be considered to be both FR and NFR. For security vs non-security requirements, the same observations as for FR/NFR hold. Finally, the adopted metrics for evaluation (precision, recall, weighted F1, accuracy) are those typically used for ML systems, so we do not foresee any major construct validity issue in this aspect. 

\paragraph{Internal Validity} As our experiments deal with software subjects, which require only limited human intervention, this ensures minimal bias. In the evaluation, we have used established and widely used annotated datasets from the literature. The only internal validity threats are somewhat inherited from the labelling performed by previous work. While the accuracy of the labelling of the PROMISE dataset has been questioned by previous work~\cite{dalpiaz2019requirements,hey2020norbert}, the dataset represents a classical benchmark, which can be used to compare our results with previous proposals. Concerning internal threats due to implementation issues, we have adopted widely used LMs. These models have been tested in other environments, thus increasing confidence in their reliability. Concerning the implementation of the ZSL approach, we have used the Transformers package in Python to retrieve the LMs from HuggingFace hub and to apply encoding for the labels and requirements representations. This package is also widely used, and we have made our code available for inspection in a Google Colab Notebook, so that the results can be replicated. Another possible threat is related to expert-curated labels. To mitigate bias in label selection, we followed a procedure of independent selection, followed by majority voting, and we reported the obtained agreements, which was moderated in all the cases. 

\paragraph{External Validity}

 Our results apply to requirements classification cases that are similar to the task considered in the paper. Different results may be observed, e.g., for the classification of requirements vs non-requirements, and the extraction of relevant content from app reviews. A main threat to external validity is due to the PROMISE dataset, as the requirements in the dataset were largely written and labelled by students, and this may not be representative of industrial requirements \citep{kurtanovic2017automatically,hey2020norbert}. We agree that the quality of the dataset can affect the performance of the LMs used in our evaluation.  
 However, the PROMISE dataset has been widely used in the RE community, as a \textit{de facto} benchmark for requirements classification, and using this dataset for research evaluation will allow RE researchers to compare their results with ours. We also make our code and data publicly available so that further replication or reproduction of our approach can be carried out. We also recognize that the lack of labelled requirements datasets has been an open challenge to using ML approaches for RE tasks \cite{binkhonain2019review}. 

\paragraph{Conclusion Validity}

To reduce the threat to our conclusion, we used statistical significance tests to compare the variance of the means within the ZSL classifiers to assess if the systems have the same effect or not. We used one-way Analysis of Variance (ANOVA) with repeated measures, and verified the results with another non-parametric significance test, the Friedman Test. From all the variance testing results, all the ZSL classifiers in all tasks are \textit{statistically significant} for $\alpha = 0.05$, 
confirming that the ZSL performance results are not due to chance.
All the statistical analysis tests are reported in the online supplementary materials\footnote{\url{https://github.com/waadalhoshan/ZSL4REQ/blob/main/StatAnalysis_ZSL4RE_results.ipynb}}.

\section{Conclusion}
\label{sec:con}

This paper reports on an extensive study of using the contextual word embedding-based zero-shot learning approach for requirements classification. The study tested this approach using \textbf{4} LMs (2 generic and 2 domain-specific), \textbf{3} groups of requirements classification tasks (Task FR/NFR, Task NFR, Task Security, and their subtasks), \textbf{19} label configurations, and \textbf{2} datasets with a total of \textbf{1020} requirements. More than \textbf{360} experiments were conducted, each based on a combination of a specific LM, a specific task, a specific label configuration, and a specific dataset. The study found: 
 
 \begin{itemize}
     \item Generic LMs perform better than domain-specific LMs under the ZSL approach.
     \item Simple label selection strategies, i.e., using original labels and expert curated labels, outperform complex strategies such as word-embedding generated labels.
 \end{itemize}

The study also found that in comparison with three previously reported supervised learning approaches for requirements classification, the performance results achieved by the best ZSL classifiers (i.e., the best combinations of the LMs and label configurations) are still lower. However, the ZSL approach is fully unsupervised that does not require any labelled dataset or training. This approach therefore has the great potential to address the problem of labelled data shortages in RE and SE.

Another advantage of the ZSL approach is that it is inherently flexible. Unlike supervised approaches that require a set of \textit{fixed classes} preassigned to the dataset, the ZSL approach can classify requirements into \textit{any unseen new classes} directed by the given labels. Consequently, the ZSL approach is suitable for requirements classification tasks facing changing classification schemes. As classification schemes change, all is required is for the ZSL approach to adopt a new set of labels, which can be defined easily, as our study shows. 


Future work will consider the following directions: 1) assess ZSL for the classification of app reviews, using existing datasets made available by previous studies (cf., Dabrowski for a complete list~\cite{dkabrowski2022analysing}); 2) explore other RE tasks to frame them as classification problems suitable for ZSL; 3) replicate current experiments with the entailment-based ZSL approach, to explore whether better performance can be achieved; 4) consider the few-shot learning approach i.e., by only using a handful of labelled examples to train the classifier, and assess to what extent the shortcomings of ZSL can be addressed by including a limited set of labelled examples; 5) evaluate the effects of different label sizes (i.e., the number of words) on the performance of ZSL.

\section*{Replication}

We shared our experimentation settings including Colab notebook and the results we obtained from all the ZSL classifiers at \url{https://github.com/waadalhoshan/ZSL4REQ}.




\section*{Acknowledgments}
We wish to thank the two reviewers for their valuable comments and suggestions, for their support and interest in this paper. We thank the Guest Editors of this special issue, Vincenzo Gervasi and Andreas Vogelsang, for their great support and interest in our paper.  Liping Zhao and Waad Alhoshan extend their appreciation to the Deanship of Scientific Research at IMSIU for funding and supporting this work through Research Partnership Program \textbf{no. RP-21-07-03}.

 \bibliographystyle{elsarticle-num} 
 
 \bibliography{bibliography}

\end{document}